\documentclass{elsart}

\usepackage{graphicx}

\usepackage{amssymb}
\usepackage{longtable}
\begin{document}

\begin{frontmatter}
\title{Measurement of the pressure dependence of air fluorescence emission  induced by electrons}

\author[Chi]{{\bf AIRFLY Collaboration}: M. Ave},
\author[CZ1]{ M. Bohacova},
\author[LNF]{ B. Buonomo},
\author[Chi]{ N. Busca},
\author[Chi]{ L. Cazon},
\author[ANL]{ S.D. Chemerisov},
\author[ANL]{ M.E. Conde},
\author[ANL]{ R.A. Crowell},
\author[AqU]{ P. Di Carlo},
\author[RomeU]{ C. Di Giulio},
\author[CZ2]{ M. Doubrava},
\author[LNF]{ A. Esposito},
\author[Sant]{ P. Facal},
\author[ANL]{ F.J. Franchini},
\author[KarlsruheU]{ J. H\"orandel},
\author[CZ1]{ M. Hrabovsky},
\author[AqU]{ M. Iarlori},
\author[ANL]{ T.E. Kasprzyk},
\author[KarlsruheU]{ B. Keilhauer},
\author[FZK1]{ H. Klages},
\author[FZK2]{ M. Kleifges},
\author[ANL]{ S. Kuhlmann},
\author[LNF]{ G. Mazzitelli},
\author[CZ1]{ L. Nozka},
\author[KarlsruheU]{ A. Obermeier},
\author[CZ1]{ M. Palatka},
\author[AqU]{ S. Petrera},
\author[RomeU]{ P. Privitera\corauthref{cor1}},
\ead{priviter@roma2.infn.it}
\author[CZ1]{ J. Ridky},
\author[AqU]{ V. Rizi},
\author[RomeU]{ G. Rodriguez},
\author[AqU]{ F. Salamida},
\author[CZ1]{ P. Schovanek},
\author[ANL]{ H. Spinka},
\author[RomeU]{ E. Strazzeri},
\author[Mun]{ A. Ulrich},
\author[ANL]{ Z.M. Yusof},
\author[CZ2]{ V. Vacek},
\author[RomeS]{ P. Valente},
\author[RomeU]{ V. Verzi},
\author[FZK1]{ T. Waldenmaier}
\footnotesize
\address[Chi]{ University of Chicago, Enrico Fermi Institute, 5640 S. Ellis Ave., Chicago, IL 60637, United States}
\address[CZ1]{Institute of Physics of the Academy of Sciences of the 
Czech Republic, Na Slovance 2, CZ-182 21 Praha 8, Czech 
Republic}
\address[LNF]{Laboratori Nazionali di Frascati dell'INFN, INFN, Sezione di Frascati, Via Enrico Fermi 40, Frascati, Rome 00044, Italy }
\address[ANL]{Argonne National Laboratory, Argonne, IL 60439 United States}
\address[AqU]{Dipartimento di Fisica dell'Universit\`{a} de l'Aquila and 
INFN, Via Vetoio, I-67010 Coppito, Aquila, Italy}
\address[RomeU]{Dipartimento di Fisica dell'Universit\`{a} di Roma Tor 
Vergata and Sezione INFN, Via della Ricerca Scientifica, I-00133 Roma, Italy}
\address[CZ2]{Czech Technical University, Technicka 4, 16607 Praha 6, Czech Republik}
\address[Sant]{ Departamento de F\'{\i}sica de Part\'{\i}culas, Campus Sur, Universidad, E-15782 Santiago de Compostela, Spain}
\address[KarlsruheU]{ Universit\"{a}t Karlsruhe (TH), Institut f\"{u}r Experimentelle Kernphysik (IEKP), Postfach 6980, D - 76128 Karlsruhe, Germany }
\address[FZK1]{Forschungszentrum Karlsruhe, Institut f\"{u}r Kernphysik, 
Postfach 3640, D - 76021 Karlsruhe, Germany}
\address[FZK2]{Forschungszentrum Karlsruhe, Institut f\"{u}r Prozessdatenverarbeitung und Elektronik, Postfach 3640, D - 76021 Karlsruhe, Germany}
\address[Mun]{Physik Department E12, Technische Universit\"{a}t Muenchen,
James Franck Str. 1, D-85748 Garching, Germany}
\address[RomeS]{Sezione INFN di Roma 1, Ple. A. Moro 2, I-00185 Roma, Italy}
\corauth[cor1]{corresponding author}

\begin{abstract}
The fluorescence detection of ultra high energy ($\gtrsim 10^{18}$ eV) cosmic rays requires a detailed knowledge of the fluorescence light emission from nitrogen molecules, which are  excited by the cosmic ray shower particles along their path in the atmosphere. 
We have made a precise measurement of the fluorescence light spectrum excited by MeV electrons in dry air. We measured the relative intensities of 34 fluorescence bands in the wavelength range from 284 to 429 nm with a high resolution spectrograph. The pressure dependence of the fluorescence spectrum was also measured from a few hPa up to atmospheric pressure. Relative intensities and collisional quenching reference pressures for bands due to transitions from a common upper level were found in agreement with theoretical expectations.  
The presence of argon in air was found to have a negligible effect on the fluorescence yield.     
We estimated that the systematic uncertainty on the cosmic ray shower energy due to the pressure dependence of the fluorescence spectrum is reduced to a level of 1\% by the AIRFLY results presented in this paper.
\end{abstract}

\end{frontmatter}

\section{Introduction}
\label{sec:intro}
     The detection of ultra high energy ($\gtrsim 10^{18}$ eV) cosmic rays using nitrogen fluorescence light emission from extensive air showers (EAS) is a well established
technique, used by the Fly's Eye \cite{flyseye}, HiRes \cite{hires}, and
Pierre Auger Observatory \cite{auger} experiments, and planned for the
Telescope Array \cite{telaray}, which is presently under construction.  It
has also been proposed for the satellite-based EUSO \cite{euso} and OWL
\cite{owl} projects. Excitation of atmospheric nitrogen by EAS charged particles induces fluorescence emission, mostly in the wavelength range between 300 to 430 nm. 
Information on the longitudinal EAS development
can be obtained by fluorescence telescopes by recording the light
intensity as a function of time and incoming direction; this
information is related to the primary cosmic ray energy and type.
However, the fluorescence light yield from EAS charged particles must be
well known at each point within the shower, and corrections applied for
atmospheric effects between the shower and the telescope for an
accurate primary energy determination. Thus, the intensities of the fluorescence bands should be measured over a range of air pressure and temperature corresponding to altitudes up to several tens of km, the typical elevation of EAS development in the atmosphere. 

A number of other experiments have made measurements of the
fluorescence light yield pertinent to EAS.  These include early
low-energy stopped-particle  results in air by Bunner \cite{bunner} and electrons in air by Davidson and
O'Neil \cite{david}.  More recently, Kakimoto {\it et al.} \cite{kakim}
obtained the light yields with three narrow and one broad band
optical filters with electrons from a radioactive source at 1.4 MeV
and from a synchrotron at 300, 650, and 1000 MeV in dry air and pure
nitrogen. Nagano  {\it et al.} \cite{nagano1} measured the light yields in
pure nitrogen and dry air with electrons of average energy 0.85 MeV
from a radioactive source, using 14 narrow-band filters.
Belz {\it et al.} \cite{belz} used 28.5
GeV electrons in pure nitrogen and dry air to measure the light
yield through a HiRes broad band optical filter ($\sim$300-400 nm) as a function of pressure. Colin {\it et al.} \cite{colin} measured the fluorescence light yield with a broad band optical filter at electron energies of 1.5 MeV, 20 GeV and 50 GeV. 
The application of these fluorescence measurements to
EAS experiments has been described recently by de Souza {\it et al.}
\cite{desouza}, Keilhauer {\it et al.} \cite{keilhauer}, and Arqueros {\it et al.} \cite{arqueros}.

The uncertainty on the fluorescence light yield is currently one of the main systematic uncertainties on the cosmic ray energy determination by EAS experiments which employ the fluorescence technique. The absolute fluorescence yield is known only at the level of 15\% and for a few electron energies. Recent spectral measurements are limited by the use of optical filters, while the early data of Bunner are based on measurements with coarse spectral resolution and large systematic uncertainties. 
  The data reported here are part of a program by the AIRFLY (AIR FLuorescence Yield)
collaboration to measure the fluorescence light yield with significantly improved precision over the electron kinetic energy range from keV to GeV using several accelerators.   This paper describes
the measurement of the pressure dependence of the relative yields of the fluorescence bands in the wavelength range 284 - 429 nm. The absolute yield, as well as the temperature and humidity dependence of the fluorescence spectrum, are currently being analysed, and will be reported elsewhere.

The relative intensities of 34 fluorescence bands over the wavelength range 284 - 429 nm were measured in dry air at 800 hPa with a high resolution spectrograph. The pressure dependence of the fluorescence spectrum was also studied from a few hPa up to atmospheric pressure. The high beam current needed for a measurement of the fluorescence spectrum  was provided by a DC beam of 3 MeV electrons.
The pressure dependence of the 337 nm band in dry air was measured in a different set-up using a narrow band optical filter and a photomultiplier tube. We used a 14 MeV electron beam for this measurement, which had the advantage of much better stability of the beam position, a small spot size and reduced multiple scattering effects compared to the 3 MeV beam. These beam characteristics were important to reduce the systematic uncertainties. We  measured the collisional quenching reference pressure of the 337 nm band, $p'_{air}$(337), by studying the ratio of fluorescence emission in nitrogen and air. With this method, we eliminated the bias from undetected light due to secondary electrons escaping the detector's field of view at low pressures.  
The collisional quenching reference pressures of the other fluorescence bands were obtained from the pressure dependence of their relative intensities, using $p'_{air}$(337) as normalization.

     This paper is organized as follows.  Section 2 presents equations
used to parameterize the pressure dependence of the fluorescence
light yield.  The experimental method and hardware are given in Section
3.  The measured fluorescence spectrum at 800 hPa in dry air, the
method used to estimate the band intensities, and a comparison to
theoretically expected intensities are discussed in Section 4.  The 337
nm pressure dependence measurements at 14 MeV are given in Section 5, as well as a
new analysis of the data which minimize the systematic uncertainties due to the spatial distribution of the fluorescence emission induced by secondary electrons. The measurement of the pressure dependence of the remaining
spectral bands is presented in Section 6. Section 7 presents applications
to EAS, and Section 8 summarizes this work.

\section{Fluorescence yield}
\label{sec:fluoreff} 
Electrons passing through air excite the nitrogen molecules, directly as well as through secondary electrons produced along the path in the gas. Nitrogen de-excitation results in a fluorescence spectrum which, in the range 300 to 430~nm, consists mainly of transitions from the so-called second positive system (2P) of molecular nitrogen $\rm{N_2}$ and the first negative system (1N) of ionised nitrogen molecules $\rm{N_2^+}$. Throughout the paper, the notations 2P($v,v'$) and 1N($v,v'$), corresponding to $C^3\Pi_u(v) \rightarrow B^3\Pi_g(v')$ and 
 $B^2\Sigma_u^+(v) \rightarrow X^2\Sigma_g^+(v')$ \cite{lofthus}, will be used to indentify the relevant electronic-vibrational transitions. Not all the excited nitrogen molecules emit fluorescence photons, since they may transfer their energy to other molecules through collision. This quenching process introduces a dependence of the fluorescence emission on the gas pressure and temperature, because the collisional rate depends on the average separation distance and velocity of the molecules.    

The process of fluorescence emission induced by
electrons in nitrogen gas can be described in terms of fluorescence
efficiency \cite{bunner} \cite{nagano1}, namely the ratio of the energy emitted by
the excited gas in fluorescence photons to the energy deposited in
the gas by the electron. The fluorescence efficiency of photons of
wavelength $\lambda$ at a given nitrogen pressure $p$ is usually
parameterized as:
\begin{equation}
\Phi_{\rm{N_2}}(\lambda,p) = \frac{\Phi^0_{\rm{N_2}}(\lambda)}{1+\frac{p}{p'_{\rm{N_2}}(\lambda)}},
\label{eq:fluoreff}
\end{equation}
where $p'_{\rm{N_2}}(\lambda)$ is the collisional quenching reference pressure, and $\Phi^0_{\rm{N_2}}(\lambda)$ is the
fluorescence efficiency in absence of collisional quenching (in the limit $p \rightarrow 0$ the distance between
nitrogen molecules becomes very large and they cannot de-excite by collisions).
Here, the {\it fluorescence yield} is defined as the ratio of the fluorescence efficiency to the
photon energy $E_\lambda$, that is the number of photons emitted by the excited gas per energy deposited by the electron:
\begin{equation}
Y_{\rm{N_2}}(\lambda,p) = \frac{Y^0_{\rm{N_2}}(\lambda)}{1+\frac{p}{p'_{\rm{N_2}}(\lambda)}},
\label{eq:fluoryield}
\end{equation}
where $Y^0_{\rm{N_2}}(\lambda)=\Phi^0_{\rm{N_2}}(\lambda)/E_\lambda$.

The quenching reference pressure for nitrogen can be written in terms of the lifetime of the excited state to decay to any lower state $\tau_0$ and the cross section for nitrogen-nitrogen collisional de-excitation $\sigma_{\rm{NN}}$ \cite{bunner}\cite{nagano1}:
\begin{equation}
 \frac{1}{p'_{\rm{N_2}}} = \frac{4 \tau_0 }{\sqrt{\pi M_{\rm{N}} kT}}\sigma_{\rm{NN}},
\label{eq:pprime}
\end{equation}
where $M_{\rm{N}}$ is the $\rm{N_2}$ molecular mass, $k$ is Boltzmann's
constant, and $T$ is temperature.

The fluorescence yield in air is given by:
\begin{equation}
Y_{air}(\lambda,p) = \frac{Y^0_{\rm{N_2}}(\lambda) f_{\rm{N_2}}}{1+\frac{p}{p'_{air}(\lambda)}},
\label{eq:fluoreffair}
\end{equation}
where $ f_{\rm{N_2}}$ is the fraction of nitrogen molecules in air (79\%). 
The quenching reference pressure in air takes into account also nitrogen-oxygen collisional de-excitation:
\begin{equation}
 \frac{1}{p'_{air}} = \frac{4 \tau_0 }{\sqrt{\pi M_{\rm{N}} kT}}\left( f_{\rm{N_2}} \sigma_{\rm{NN}} + f_{\rm{O_2}} \sigma_{\rm{NO}}\sqrt{\frac{M_{\rm{N}}+M_{\rm{O}}}{2M_{\rm{O}}}}\right) = \frac{ f_{\rm{N_2}}}{p'_{\rm{N_2}}}+\frac{ f_{\rm{O_2}}}{p'_{\rm{O_2}}} ,
\label{eq:pprimeair}
\end{equation}
where  $ f_{\rm{O_2}}$ is the fraction of oxygen molecules in air (21\%), $M_{\rm{O}}$ is
the $\rm{O_2}$ molecular mass and  $\sigma_{\rm{NO}}$ is  the cross section for
nitrogen-oxygen collisional de-excitation.  
In this paper, we will not
interpret the measurements in terms of collisional cross sections, but rather
use the quenching reference pressure for each band as a phenomenological
description of the data.
Note that the fluorescence yield process finds in Eqs.~(\ref{eq:fluoreffair})~-~(\ref{eq:pprimeair}) its simplest description. Mechanisms which may modify  Eqs.~(\ref{eq:fluoreffair})-(\ref{eq:pprimeair}) have been discussed in the literature (see for ex. \cite{keilhauer}\cite{mitchell}\cite{calo}). On the other hand, it will be shown in the following Sections that the ansatz described by Eqs. (\ref{eq:fluoreffair})-(\ref{eq:pprimeair}) does indeed provide a good description of our data.  

     The typical arrangement in fluorescence experiments involves a photon
detector, e.g. a photomultiplier, viewing a portion of gas volume ($\rm{N_2}$ or
air) along the electron path.  The detected signal is thus proportional to
the number of fluorescence photons $N_{\lambda}^{gas}$ emitted in the
detector field of view:
\begin{equation}
 N_{\lambda}^{gas} = E_{dep}^{gas} Y_{gas}(\lambda,p),
\label{eq:nlambda}
\end{equation}
where $E_{dep}^{gas}$ is the energy deposited by the electron in the
gas volume viewed by the photon detector.  The energy
$E_{dep}^{gas}$ depends on the specific geometry of the experiment,
and can be estimated by a detailed Monte Carlo simulation, including
effects like multiple scattering and secondary electrons.
Nevertheless, for a given experimental arrangement, we can write:
\begin{equation}
 E_{dep}^{gas} = \left( \frac{dE}{dX} \right)_{gas} \rho_{gas} D_{gas}(p) ,
\label{eq:edep}
\end{equation}
where $(dE/dX)_{gas}$ is the collisional energy loss, $\rho_{gas}$ is the gas density and $D_{gas}(p)$ is an effective length which takes into account the specific experimental arrangement.
Notice that a pressure dependence has been included in  $D_{gas}(p)$. In fact, as the gas pressure goes down, an increasing fraction of secondary electrons escapes the detector field of view, and correspondingly part of the fluorescence emission is not detected.
The net effect is a reduction of the effective length  $D_{gas}$, which is thus expected to be a decreasing function of $p$.

The number of detected fluorescence photons can then be written:
\begin{equation}
 N_{\lambda}^{gas} =  \left( \frac{dE}{dX} \right)_{gas} \frac{p}{R_{gas}T}  D_{gas}(p) Y_{gas}(\lambda,p),
\label{eq:nlambdaf}
\end{equation}
 where $R_{gas}$ is the specific gas constant, $T$ is the gas temperature
and the gas equation of state has been used.

\section{Experimental method}
\label{sec:expmethod}

     Measurements were performed at two Argonne National Laboratory
accelerators.  Spectra from 284 - 429 nm were recorded with 3~MeV
electrons in various gases and pressures using the Chemistry Division's
Van de Graaff (VdG) electron accelerator \cite{VdG} \cite{VdGsor}.
The relative light yield of the 337 nm band was measured at 14~MeV
at the Argonne Wakefield Accelerator (AWA) Facility \cite{AWA}
in dry air and pure nitrogen as a function of pressure.  This relative
light yield, combined with the measured spectra, allowed the pressure
dependence of the fluorescence bands to be derived.

\subsection{Van de Graaff Measurements}
\label{sec:vdgsec}
\begin{figure}
\begin{center}
\includegraphics*[width=12cm]{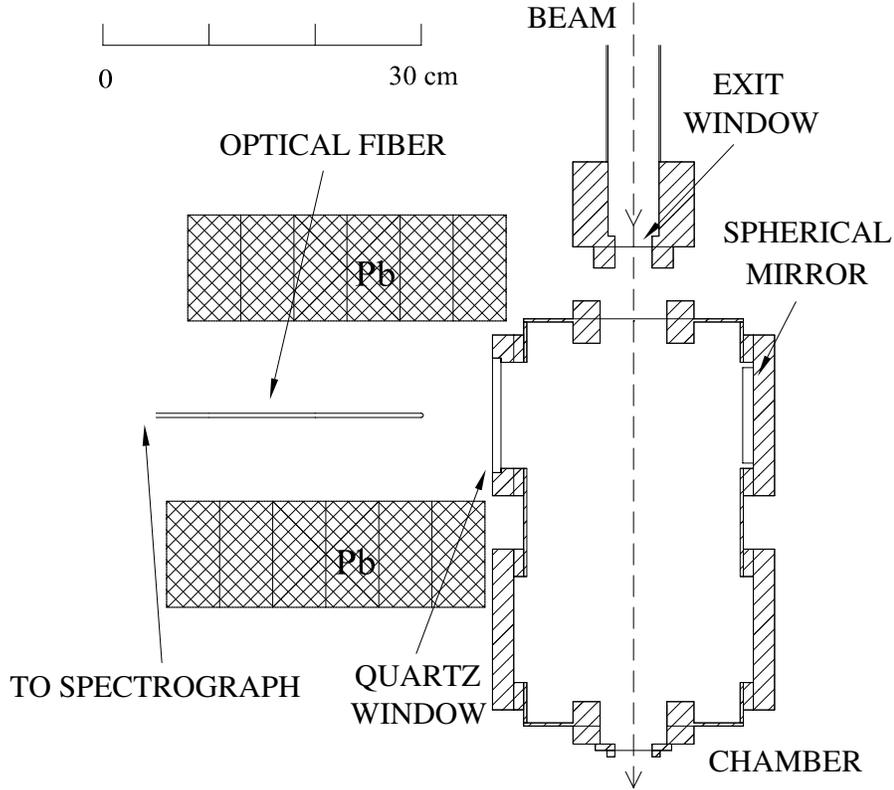}
\end{center}
\caption{Layout of the measurements at the Van de Graaff, including the
  exit window from the accelerator vacuum, the pressure chamber,
  spherical mirror, quartz window, and optical fiber.}
\label{fig:VdG}
\end{figure}

     The Van de Graaff accelerator is capable of accelerating electrons
and protons
to kinetic energies from 0.5 to 3 MeV,  with beam currents in excess
of 10 $\mu A$.  The VdG can operate in DC current or pulsed mode.
For the spectrum measurements it was operated in the DC current mode
with typical beam currents of $\sim 10$ $\mu A$, and nominal beam
kinetic energy of 3.0 MeV. After exiting the VdG, the electron beam
was bent $30^{\circ}$ in an electromagnet and was focused near the
exit window 2.00 m from the magnet center. The electrons left the
accelerator vacuum through a 35 mm diameter, 0.152 mm thick
dura-aluminum window; see Fig. \ref{fig:VdG}.  An additional quartz
window was placed downstream of the aluminum window during beam
tuning.  The beam spot image from the quartz was reflected with a
mirror to a camera.  The beam spot size was typically 6 mm
diameter, and a side-to-side beam motion of approximately 5 mm was
observed due to small ($< 1 \%$) variations in the VdG energy on
time scales of seconds. The quartz window and mirror were removed
during measurements of the spectra.

     The layout for the Van de Graaff measurements is shown in
Fig. \ref{fig:VdG}.  The electrons entered the pressure chamber after
6 cm of air.  Light produced in the gas was focused by an aluminum
spherical mirror onto the end of a 10 m long, 1.5 mm diameter pure
silica core optical fiber, which brought the light to a spectrograph.
The optical fiber was placed outside the pressure chamber, and light
reached the fiber end passing through a quartz window.  The spectrograph
was located behind a concrete block wall and had additional lead
shielding to protect it from the radiation produced by the VdG.

\subsection{Argonne Wakefield Accelerator Measurements}
\label{sec:awasec}
\begin{figure}
\begin{center}
\includegraphics*[width=14cm]{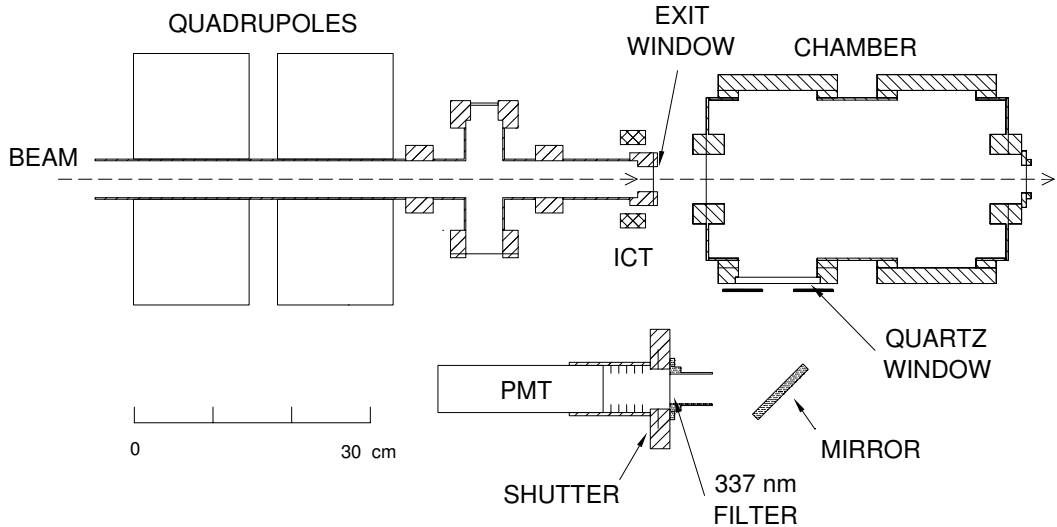}
\end{center}
\caption{Layout of the measurements at the Argonne Wakefield Accelerator, 
  including the exit window from the accelerator vacuum, ICT, the pressure 
  chamber, quartz window, mirror, 337 nm filter, mechanical shutter, and 
  photomultiplier.  The last two quadrupoles are also shown, but not the 
  third quadrupole or the adjacent dipole.  Considerable lead 
  shielding of the photomultiplier and mirror were also omitted for 
  clarity.}
\label{fig:AWA}
\end{figure}

     The Argonne Wakefield Accelerator can accelerate electrons to
kinetic energies
from 3 MeV to 16 MeV. It was operated in the pulsed mode at 5
Hz, with bunches of maximum charge of $\sim 1$ nC and length $\sim
15$ ps (FWHM). A new set of three quadrupole magnets and two Vernier
steering dipole magnets was added for this experiment to focus and
steer the beam. The nominal beam kinetic energy was 14.0 MeV, with
an estimated energy spread of $\pm 0.3$~MeV.  The electrons exited
the accelerator vacuum through a 32 mm diameter, 0.13~mm thick
beryllium window; see Fig. \ref{fig:AWA}.  During beam tuning, the beam spot
image from a quartz plate placed at $45^{\circ}$ to the beam
direction near the exit window was viewed with a camera.  The beam
spot size was typically 5 mm diameter, with negligible beam
motion. The quartz plate was removed during the fluorescence
measurements.

     The beam intensity was monitored with an integrating current transformer
(ICT,\cite{ict}), immediately before the beam exit flange.  The signal
from the ICT was integrated, digitized, and recorded for each beam bunch.
The ICT was calibrated by the manufacturer to give an output charge 0.025
times the input charge.

     The layout for the AWA measurements is shown in Fig. \ref{fig:AWA}, 
including the last two quadrupoles and ICT.  Light produced in the
gas propagated out of the chamber through a quartz window, mechanical
shutter, and 337 nm filter, and was reflected by a mirror to a
photomultiplier.  The shutter could be closed remotely to allow
measurements of background.  A Hamamatsu H7195P photomultiplier \cite{pmt} 
was used.  A second photomultiplier of the same
type was used to monitor backgrounds.  Both were surrounded by
considerable lead shielding to reduce beam-related backgrounds.  The
first photomultiplier, mirror, filter, and shutter are also shown in 
Fig. \ref{fig:AWA}.

     A VME data acquisition system was used.  The
signals from the photomultipliers and the ICT were integrated by a
Lecroy model 1182 charge integrating ADC.  The
accelerator timing signal was used to produce the integrating gate
of 200 ns width.  Signals were recorded for each electron bunch
passage in the pressure chamber.

\subsection{Pressure Chamber}
\label{sec:chambersec}
     The pressure chamber was constructed of an aluminum tube with
various flanges welded to it for windows, gauges, gas inlet, and
pump-out. The aluminum tube had an inner diameter of 201 mm, length
378 mm, and wall thickness 3 mm.  The exit window of 0.1 mm thick
aluminum was bolted to the one end.  The entrance window of 0.50 mm
thick beryllium, $35 \times 55$ mm, was diffusion bonded to a
Conflat flange.  All vacuum seals were made with O-rings.  A top
view of the pressure chamber is shown in Fig. \ref{fig:VdG}.

     A remotely-controlled gas handling and vacuum system was used with
the pressure chamber.  A dry scroll vacuum pump \cite{pump}
was used to reduce the pressure and evacuate the chamber.  The
chamber pressure was measured \cite{gage} at the pump-out
port.  Three types of high
purity dry gases were used: 99.9995\% pure nitrogen, a mixture with
argon (78.0\% nitrogen, 21.0\% oxygen, 1.0\% argon), and a mixture
without argon (79\% nitrogen, 21\% oxygen). 

\subsection{Spectrograph}
\label{sec:spectrographsec}
     Air fluorescence spectra were recorded by an Oriel MS257$^{TM}$
spectrograph \cite{spec}.  The main characteristics of the instrument
include an
asymmetrical Czerny-Turner design, F number equal to 3.9, and focal
length of 220 mm.  A holographic grating with 1200 lines/mm and
blaze wavelength of 250 nm was used.  The light from the grating was
collected by a 1024x255 CCD pixel array (Andor DV420 BU2
\cite{ccd}).  The CCD camera, of 26x26 $\mu$m pixel size, was
back-illuminated and had a large quantum efficiency ($\sim$50\%) in
the 300 to 400~nm wavelength range.  The linear dispersion of spectrograph, with the grating used, is 3.2 nm/mm, that gives a wavelength window of about 85 nm.   
This means that the air fluorescence spectrum, which extends over more than 100~nm, did not fit in a single wavelength window of the spectrograph.
Thus, in a fluorescence run the spectrograph collected data in
sequence, first in the range 284-369 nm and then in the range
344-429~nm.  For each wavelength range, 50 spectra of 1 second
integration time were taken.  The spectrograph operation and data
acquisition was fully automated.

The spectral bands emitted by a mercury pencil lamp (Oriel no.
60635 \cite{spec}) in the region 290 to 410 nm were used to calibrate
the spectral response of the spectrograph.  The calibration was then
refined by including the positions of the nitrogen emission
bands as seen in the air fluorescence spectrum measurement.

A quartz tungsten halogen lamp (Oriel no. 63350 \cite{spec}) was used
to calibrate the relative spectrograph sensitivity as a function of
wavelength. The spectral irradiance of the lamp is traceable to NIST
\cite{NIST}, and its uncertainty amounts to 2.5\% in the
wavelength range of interest for the air fluorescence measurement.

The spectrograph calibration was performed on the beam line, between
air fluorescence measurements. A flange on the top of the pressure
chamber allowed the insertion of the calibration lamp along the beam
path. In order to take into account the spatial distribution of the
air fluorescence emission in the experiment, the calibration lamp
was placed in different positions along the beam axis and the
recorded lamp spectra were appropriately combined. By comparison
with the known lamp spectrum, calibration factors for the relative
intensity as a function of wavelength were determined. The two
spectrograph ranges used for the air fluorescence measurement,
284-369 nm and 344-429 nm, were independently calibrated. Their
relative normalization, after the calibration procedure, was checked
by comparing the intensities of the mercury pencil lamp lines
emitted in the overlap region of the two spectrograph's ranges. A
2\% difference in the measured intensities was found, which was
taken as an estimate of the systematic uncertainty on the
spectrograph relative calibration procedure.

\section{Measurement of air fluorescence spectrum}
\label{sec:relintensity}

\begin{figure}
\begin{center}
\includegraphics*[width=14.cm]{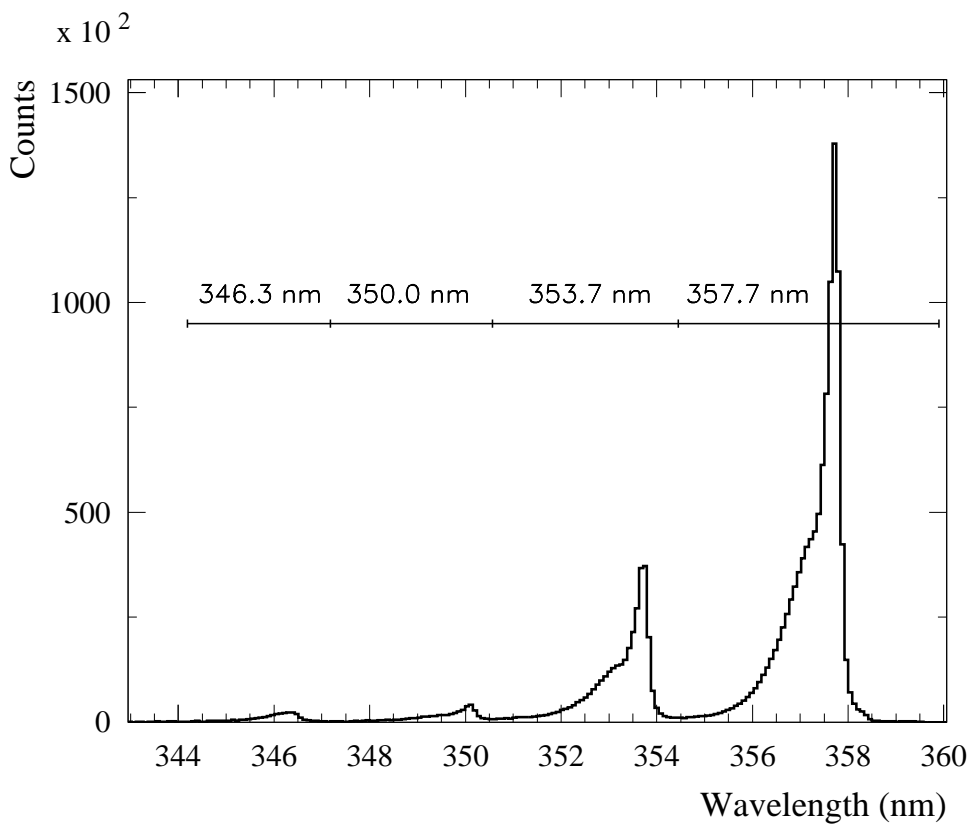}
\end{center}
\caption{Example of band integrals. Horizontal lines indicate the wavelength ranges used for the integration of the band fluorescence signal. The main fluorescence band corresponding to each interval is also given.}
\label{fig:spectzoom}
\end{figure}

     The fluorescence spectrum in air-like mixtures has been measured
at the Van de Graaff.  Details of the beam and apparatus were given
in Section \ref{sec:vdgsec}. Measurements were performed at a gas
pressure of 800 hPa and temperature of 293 K. A blow-up of the
spectrum in the wavelength region between 343 nm and 360 nm is shown
in Fig. \ref{fig:spectzoom}.  Notice that the observed lineshapes
show an asymmetric tail towards smaller wavelengths, due to the
convolution of the rotational levels with the finite resolution of
the spectrograph.  Thus, in general, the integral counts associated
with a given band can have contributions also from nearby bands.
Rather than attempting to disentangle individual band integrals,
which would result in additional systematic uncertainties, the band
intensities were estimated by integrating the CCD counts in a
wavelength interval around each band.  An example of the procedure
is shown in Fig. \ref{fig:spectzoom}. Wavelength intervals up to
several nm were used, depending on the distance of nearby bands and
the amount of rotational contribution.  In any case, most of the
signal of each band is contained in only 2 nm.  In the following,
the wavelength of the main band is used to identify the
corresponding wavelength interval.  The measurement of the light
intensity in wavelength intervals of nm size is well suited for the
purpose of calibration of the cosmic ray fluorescence detection
technique, and assigning the full intensity to the main band of the
wavelength interval will result in a negligible systematic
uncertainty.

The measured fluorescence spectrum contained a small level of background, which was estimated
in selected wavelength regions where fluorescence bands were not present. For each of these
background regions, the average number of background counts per wavelength bin was determined.
In order to estimate the background level under the fluorescence bands, a linear interpolation
was performed between the measurements of consecutive background wavelength regions. The procedure
gave an estimate of the average number of background counts in each wavelength bin over the full
wavelength region. The background level was found to be approximately constant. The amount of
background under each fluorescence band was determined by summing the estimated average number of
background counts for each bin of the wavelength interval associated to the band. The band intensity
was then determined by subtracting the corresponding background.

     Several fluorescence bands (346.3, 350.0, 353.7, 357.7, 366.1, and 367.2
nm) were common to the 284-369 nm and the 344-429 nm spectrograph's ranges.
The measured intensity of these bands was used to normalize the two
spectrograph's ranges.  In fact, any change in the beam intensity or position
during the short time (30 s) needed by the spectrograph to move from the
first to the second wavelength range would result in a difference in the
measured intensity of the common bands.  The intensity of the common bands
was found to be 2\% lower in the 344-429 nm range, and the intensity of all
measured bands in this range were corrected correspondingly.  The smallness
of the correction illustrates the good stability of the beam and the
reliability of the relative calibration of the spectrum.

\begin{figure}
\begin{center}
\includegraphics*[width=21cm,angle=90]{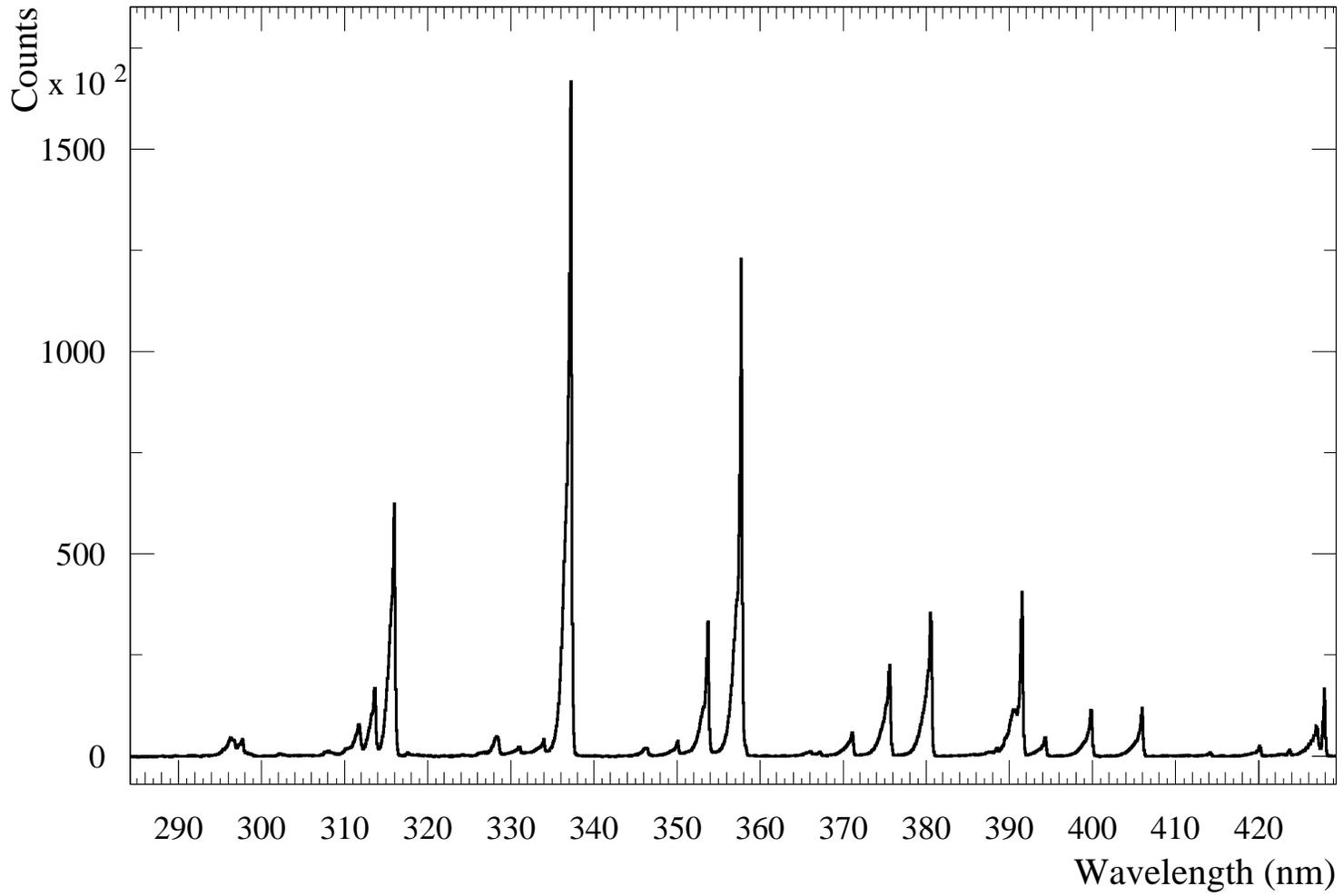}
\end{center}
\caption{Measured fluorescence spectrum in dry air at 800 hPa and 293 K.}
\label{fig:spectrum}
\end{figure}

The measured spectrum with the 78\% N$_2$ - 21\% O$_2$ - 1\% Ar
mixture is shown in Fig. \ref{fig:spectrum}.
\begin{table}[hp]
\centering
\begin{tabular}{|c|c|c||c|c|c|}
 \hline
$\lambda$ (nm)  & $\lambda$ interval (nm) &  $I_{\lambda}$ (\%) & $\lambda$ (nm) & $\lambda$ interval (nm) & $I_{\lambda}$ (\%) \\
\hline
   296.2  &292.5-297.0 &$  5.16   \pm 0.29  $ & 366.1  &363.2-366.4 &$  1.13   \pm 0.08  $\\
   297.7  &297.0-299.6 &$  2.77   \pm 0.13  $ & 367.2  &366.4-367.5 &$  0.54   \pm 0.04  $ \\
   302.0  &301.5-303.3 &$  0.41   \pm 0.06  $ & 371.1  &367.6-371.7 &$  4.97   \pm 0.22  $ \\
   308.0  &306.8-309.3 &$  1.44   \pm 0.10  $ & 375.6  &371.7-376.3 &$ 17.87   \pm 0.63  $  \\
   311.7  &309.3-312.3 &$  7.24   \pm 0.27  $ & 380.5  &376.3-381.4 &$ 27.2   \pm 1.0  $ \\
   313.6  &312.3-314.1 &$ 11.05   \pm 0.41  $ & 385.8  &383.0-386.0 &$  0.50   \pm 0.08  $\\
   315.9  &314.1-316.7 &$ 39.3   \pm 1.4  $  & 387.7  &386.0-388.0 &$  1.17   \pm 0.06  $  \\
   317.6  &317.0-318.4 &$  0.46   \pm 0.06  $ & 388.5  &388.0-388.7 &$  0.83   \pm 0.04  $ \\
   326.8  &325.6-327.1 &$  0.80   \pm 0.08  $ & 391.4  &388.7-392.1 &$ 28.0   \pm 1.0   $ \\
   328.5  &327.1-329.0 &$  3.80   \pm 0.14  $ & 394.3  &392.1-394.9 &$  3.36   \pm 0.15  $ \\
   330.9  &329.0-331.3 &$  2.15   \pm 0.12  $ & 399.8  &394.9-400.5 &$  8.38   \pm 0.29  $ \\
   333.9  &331.3-334.3 &$  4.02   \pm 0.18  $ & 405.0  &400.5-406.6 &$  8.07   \pm 0.29  $ \\
   337.1  &334.3-338.4 &$100.00     $         & 414.1  &412.5-414.4 &$  0.49   \pm 0.07  $ \\
   346.3  &344.2-347.2 &$  1.74   \pm 0.11  $ & 420.0  &416.6-420.6 &$  1.75   \pm 0.10  $ \\
   350.0  &347.2-350.6 &$  2.79   \pm 0.11  $ & 423.6  &420.7-424.0 &$  1.04   \pm 0.11  $ \\
   353.7  &350.6-354.4 &$ 21.35   \pm 0.76  $ & 427.0  &424.0-427.4 &$  7.08   \pm 0.28  $  \\
   357.7  &354.4-359.9 &$ 67.4   \pm 2.4   $ & 427.8  &427.4-428.6 &$  4.94   \pm 0.19  $  \\
\hline
\end{tabular}
\vskip 0.1truein
\caption{Measured fluorescence band intensities in dry air at 800 hPa pressure
   and 293 K temperature.  The intensity of the 337 nm band was used for
   normalization. The wavelength intervals used for the signal integration are also reported.}
\label{tab:relspectrum}.
\end{table}
The measured intensities, $I_{\lambda}$, relative to the intensity
of the 337~nm band, of 34 fluorescence bands are reported in Table \ref{tab:relspectrum}, together with the corresponding wavelength intervals used for the signal integration.  The fluorescence spectrum was measured ten times, and from the observed r.m.s.
of the ten measurements of each band intensity, uncertainties were
estimated.  To obtain the total uncertainty quoted in Table
\ref{tab:relspectrum}, a systematic uncertainty of 3.5\%, assigned
to the calibration of the relative spectrograph sensitivity, was summed
in quadrature.  The systematic uncertainty of the relative spectrograph
sensitivity accounts for the uncertainty on the absolute calibration of
the lamp emission (2.5\%), the difference in the  mercury pencil lamp
line intensities measured in the two wavelength ranges of the
spectrograph (2\%), and the uncertainty associated with the spatial
distribution of the fluorescence light source (1.5\%).  The fluorescence
light is in fact emitted along the beam axis, and the calibration
procedure of relative spectrograph sensitivity (cf. Section \ref{sec:spectrographsec})
derived calibration factors from measurements with the calibration lamp
placed in different positions along the beam axis.  If calibration factors
obtained only from a measurement with the calibration lamp placed at the
optics center are used, the relative fluorescence band intensities changed
by at most 3\%.  Half of this shift  was conservatively taken as an
estimate of the associated systematic uncertainty.

Several checks were performed. The linearity of the fluorescence emission with beam currents from 0.2 to 14 $\mu$A was verified.
Possible systematic effects due to the beam position were investigated. The fluorescence spectrum was measured with the beam moved $\pm$1 cm in the directions transverse to the nominal beam axis. No difference beyond the statistical uncertainty in the relative intensities of the fluorescence bands was found.
Several models for background evaluation were tested, which always resulted in changes of the relative band intensities within the quoted uncertainties.
A measurement of the fluorescence spectrum in pure nitrogen gas was performed,
which showed that all the observed bands in the air fluorescence spectrum are
associated with nitrogen excitation.

\begin{figure}
\begin{center}
\includegraphics*[width=13cm]{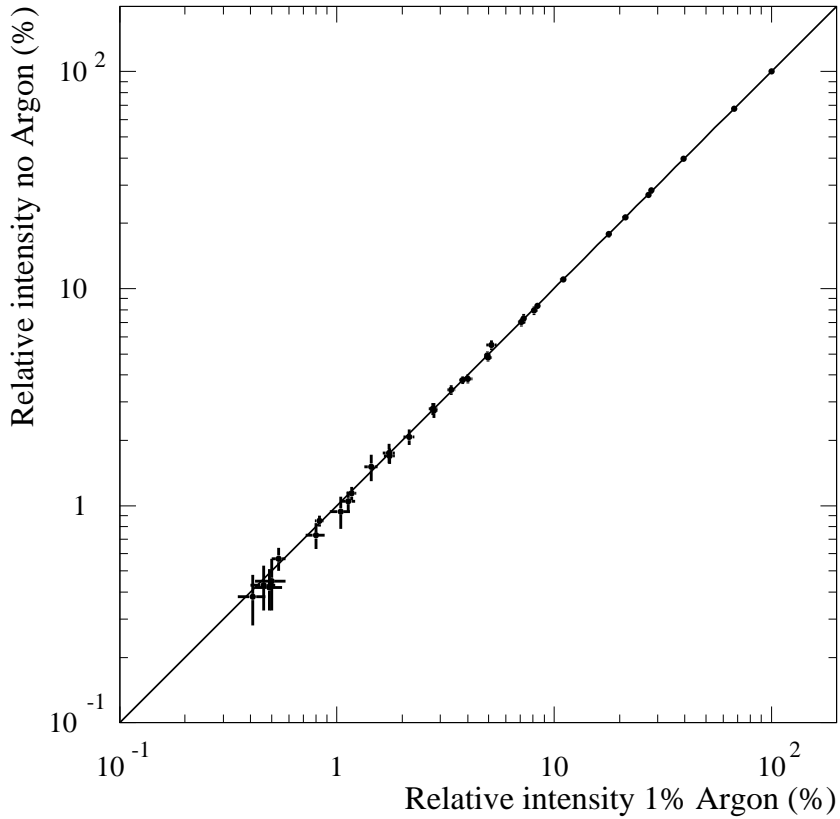}
\end{center}
\caption{Effect of argon on the air fluorescence yield. Relative intensities of the 34 fluorescence bands measured in a 79\% N$_2$ - 21\% O$_2$ mixture vs the corresponding intensities measured in the 78\% N$_2$ - 21\% O$_2$ - 1\% Ar mixture. A linear fit of the data is also shown, which indicates that the yield in the two gas mixtures is the same to a high level of accuracy.}
\label{fig:spectrumUZAr}
\end{figure}

     In order to assess the relevance of argon on air fluorescence,
the fluorescence spectrum emitted by a 79\% N$_2$ - 21\% O$_2$ gas
mixture was measured at the same pressure and temperature
conditions as the mixture with argon. Fig. \ref{fig:spectrumUZAr} shows the correlation of the
relative intensities of the 34 fluorescence bands measured with the 79\% N$_2$ - 21\% O$_2$ and the 78\% N$_2$ - 21\% O$_2$ - 1\% Ar mixture. A linear fit yielded an intercept consistent
with zero and a slope consistent with unity, within a few per mil,
which indicates that the two spectra are compatible to a high degree
of accuracy. On the basis of this observation, the effect of argon
on air fluorescence at atmospheric pressure is completely negligible.

     The 2P and 1N bands of nitrogen emission can be easily identified in
our measured fluorescence spectrum. A group of weak bands (302.0, 308.0, 317.6, 346.3, 366.1, 387.7 nm) is consistent with the Gaydon-Herman (GH) bands \cite{lofthus} \cite{filippelli}. The GH(0,1-3) bands at 296.7, 311.9 and 328.3 nm are also present, with a large overlap with the 2P(3,1-3) bands at 296.2, 311.7 and 328.5 nm. The excitation of the Gaydon-Herman bands by electrons in air at atmospheric pressure has already been observed \cite{eckstrom}.    
 Within a vibrational band system, the relative
intensities are expected to be equal to the ratio of the
corresponding Einstein coefficients, and can be calculated
\cite{gilmore} \cite{fons}.  However, the intensities reported in Table
\ref{tab:relspectrum} cannot be directly compared with the
theoretical expectations, due to the contamination from nearby
bands.  A simple extrapolation of the band's spectral shape was used
to estimate the contamination, and intensities were corrected
accordingly.  The corresponding relative intensities for the
2P(0,i), 2P(1,i), 2P(2,i), and 1N(0,i) bands are reported in Table
\ref{tab:einstein}, together with the theoretical predictions
\cite{gilmore} \cite{fons}.  Only bands where the contamination
could be reliably estimated are included in  Table
\ref{tab:einstein}.  The measured relative intensities are close to
the theoretical expectations and this represents an independent cross-check of the relative spectrograph sensitivity calibration.
\begin{table}[h]
\centering
\begin{tabular}{|c|c|c|c||c|c|c|c|}
 \hline
Band & $\lambda$ & Measured  & Theory & Band & $\lambda$& Measured & Theory\\
   & (nm) & (\%) & (\%)& & (nm) & (\%) & (\%)\\
\hline
2P(0,0) &   337.1  & 100. & 100.&  2P(2,1) &   313.6  & 100. & 100.\\
2P(0,1) &   357.7  & 67.4 & 67.5 & 2P(2,2) &   330.9  & 7.9 & 7.3\\
2P(0,2) &   380.5  & 26.5 & 27.2 & 2P(2,3) &   350.0  & 16.1 & 16.9\\
2P(0,3) &  405.0  & 7.9 & 8.4 & 2P(2,4) &   371.1  & 38.7 & 40.0 \\
\cline{1-4}
2P(1,0) &  315.9  & 100. & 100.& 2P(2,5) &   394.3  & 29.6 & 31.1\\
2P(1,1) &  333.9  & 4.0 & 4.9 & 2P(2,6) &   420.0  & 15.5 & 15.5 \\
\cline{5-8}
2P(1,2) &  353.7  & 47.8 & 46.6 & 1N(0,0) &   391.4  & 100. & 100\\
2P(1,3) &   375.6  & 41.9 & 41.4 &1N(0,1) &   427.8  & 31.8 & 32.5 \\
\cline{5-8}
2P(1,4) &   399.8  & 20.8 & 20.4  \\
2P(1,5) &   427.0  & 9.9 & 7.5  \\
\cline{1-4}

\end{tabular}
\vskip 0.1truein
\caption{Measured relative intensities within bands and theoretical expectations based
  on the ratio of Einstein coefficients.}
\label{tab:einstein}.
\end{table}

\section{Pressure dependence of the 337 nm band}
\label{sec:pres337}

\begin{figure}
\begin{center}
\includegraphics*[width=14.cm]{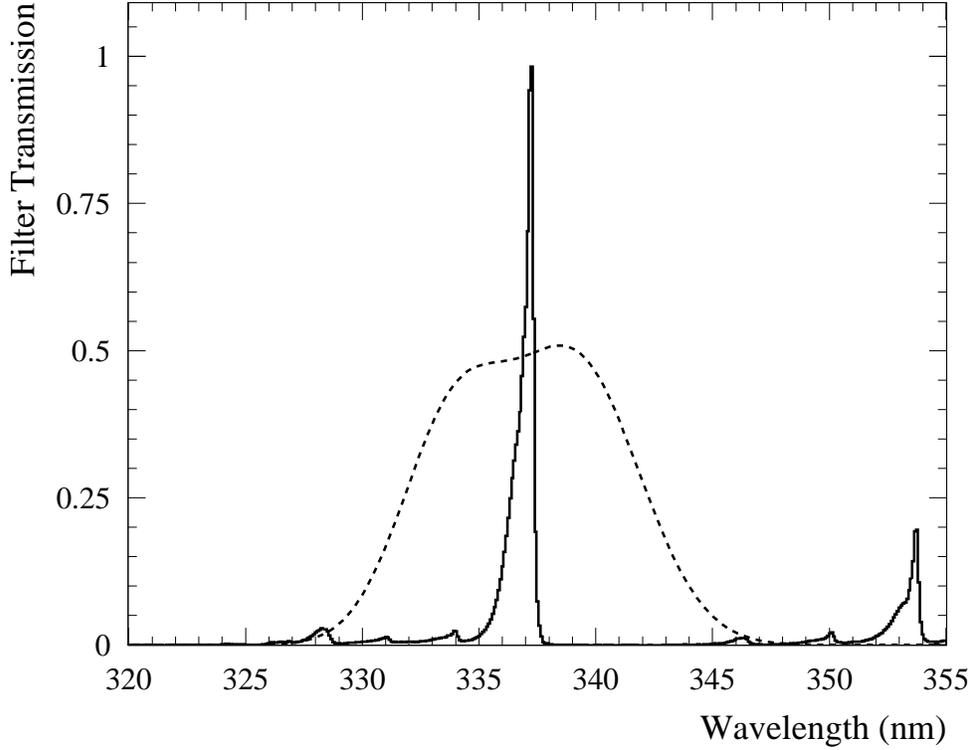}
\end{center}
\caption{Transmission of the interference filter used to select the 337 nm band (dashed line). The measured fluorescence spectrum in arbitrary units is shown as a full line.}
\label{fig:filter}
\end{figure}

A precise measurement of the pressure dependence of the 337 nm band
has been performed at the AWA facility. The experimental set-up has
been described in Section \ref{sec:awasec}. A narrow band
interference filter centered at 337 nm was used for this
measurement. The filter transmission curve is shown in Fig.
\ref{fig:filter}, together with the fluorescence emission bands as
measured by the spectrograph (Section \ref{sec:relintensity}). The filter transmission was measured by three different groups of the AIRFLY collaboration, which yielded consistent results at the percent level. The filter selects mainly
the 337 nm band, the contamination from neighbouring bands being
only 1.7\%.

\begin{figure}
\begin{center}
\includegraphics*[width=14.cm]{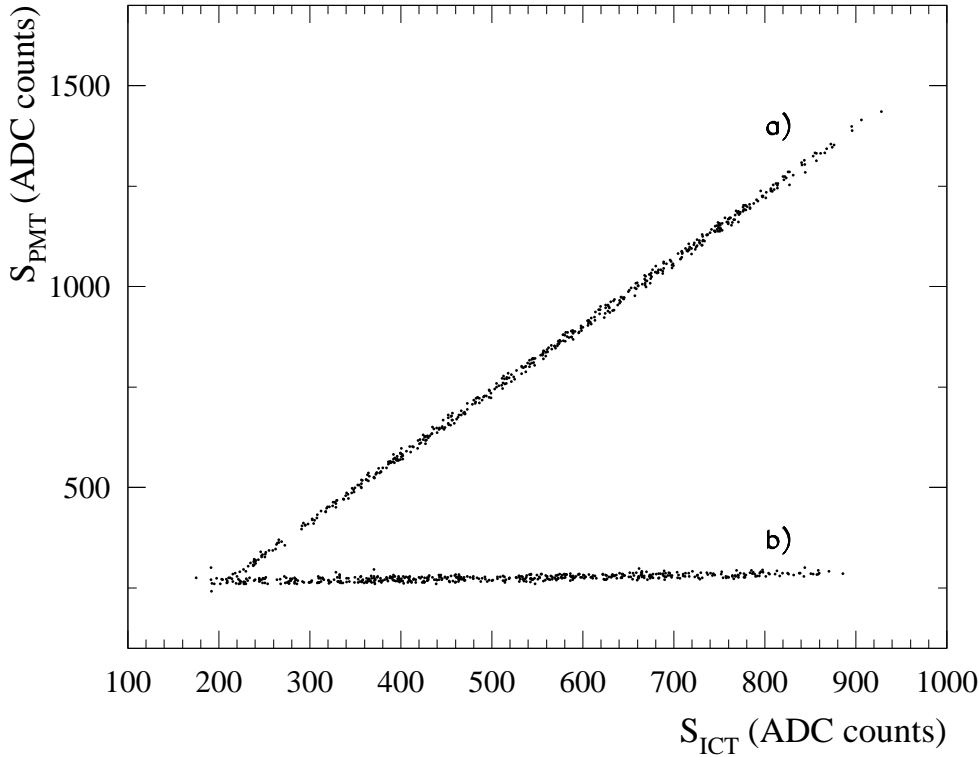}
\end{center}
\caption{ Correlation between the photomultiplier signal and the beam ICT signal: a) fluorescence run in pure nitrogen at 180 hPa, b) background run.}
\label{fig:correl}
\end{figure}

 In order to check linearity and possible pedestal changes within a run, the electron beam was
 operated in a mode allowing the bunch charge to fluctuate over a wide range. An example of
 correlation of the photomultiplier signal $S_{PMT}$ and the ICT signal $S_{ICT}$ is shown in
 Fig. \ref{fig:correl}, for a run at 180 hPa in pure nitrogen. A linear fit $S_{PMT}=S_{FL} \cdot S_{ICT}+b$ to
 the data was performed, and the fitted slope $S_{FL}$ was taken as an estimator of the fluorescence signal.
 The same procedure, applied on data taken with the shutter closed in front of the photomultiplier tube,
 allowed an estimate of the background level. A typical background correlation to the beam intensity is
 also shown in  Fig. \ref{fig:correl}.  Several background runs were taken during the pressure scans. In addition,
 during the fluorescence runs, the background was monitored by a second photomultiplier, and found to be stable
 within its statistical uncertainty. The background slope was subtracted from the signal slope, and in the
 following $S_{FL}$ will refer to the background subtracted signal.

Measurements were performed with dry air-like mixtures and pure nitrogen.
The gas pressure in the fluorescence chamber was varied in the range from 2 to 1000 hPa, and for each pressure $p$ a value $S_{FL}(p)$ was measured. The gas temperature was 293 K. Only a couple of minutes for each pressure point was necessary to collect enough statistics, thus minimizing the possibility of systematic changes of the beam and the gas condition, as well as the detector response.

The measured fluorescence signal $S_{FL}^{gas}(p)$ in nitrogen or air 
is proportional to (cf. Eq. (\ref{eq:nlambdaf})):
\begin{equation}
S_{FL}^{gas}(p) \propto N_{337}^{gas} =  \left( \frac{dE}{dX} \right)_{gas} \frac{p}{R_{gas}T} D_{gas}(p) \frac{Y^0_{\rm{N_2}}(337) f_{\rm{N_2}}}{1+\frac{p}{p'_{gas}(337)}},
\label{eq:signal}
\end{equation}
where  $f_{\rm{N_2}}=1$ for pure nitrogen.
Notice that the fluorescence emission is essentially due to excitation by
secondary electrons produced in the gas by the beam electrons
\cite{secondary} \cite{fernando}.  As the gas pressure goes down, an increasing fraction
of secondary electrons does not stop in the photomultiplier field of
view, and correspondingly part of the fluorescence emission is not
detected. Neglecting the losses due to the secondary
electrons escaping the field of view would cause an overestimation of the 
quenching reference pressure $p'$ \cite{fernando}.

In order to avoid this bias, we have studied the ratio of the 337 nm
fluorescence signal in pure nitrogen to the signal in air,
$r_{\rm{N_2}}=S^{\rm{N_2}}_{FL}(p)/S^{air}_{FL}(p)$. The effective length
$D(p)$ is expected to cancel in the ratio, since secondary electron
interactions are similar in nitrogen and air thanks to their
close molecular masses.  This was verified with a detailed GEANT4 \cite{geant4} simulation of the experiment. 
The pressure dependence of the fluorescence signal ratio thus
becomes:
\begin{equation}
r_{\rm{N_2}} = \frac{ \left(\frac{dE}{dX} \right)_{\rm{N_2}}/R_{\rm{N_2}}}{\left( \frac{dE}{dX} \right)_{air}f_{\rm{N_2}}/R_{air}}
\cdot \frac{1+p \left(\frac{f_{\rm{N_2}}}{p'_{\rm{N_2}}(337)}+\frac{f_{\rm{O_2}}}{p'_{\rm{O_2}}(337)}\right)
}{1+\frac{p}{p'_{\rm{N_2}}(337)}},
\label{eq:rn2}
\end{equation}
where $f_{\rm{N_2}}$ and $f_{\rm{O_2}}$ are the fraction of nitrogen and
oxygen molecules in the air mixture. The collisional energy loss $(dE/dX)$ in
nitrogen and air at 14 MeV was calculated with the method described in \cite{estar}.

\begin{figure}
\begin{center}
\includegraphics*[width=14.cm]{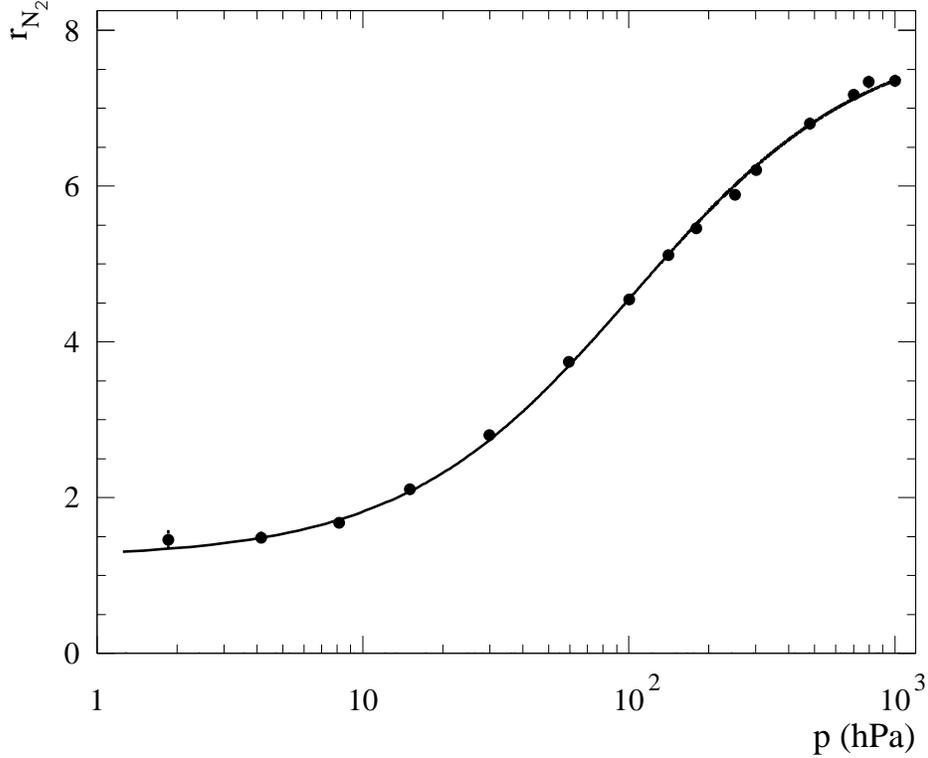}
\end{center}
\caption{Pressure dependence of the ratio of the 337 nm band
   signal in nitrogen to the signal in dry air.  The full line is the
   result of a fit to the full pressure range described in the text.}
\label{fig:ratioarg}
\end{figure}

The measured pressure dependence of the ratio  $r_{\rm{N_2}}$ of the 337
nm band is shown in Fig. \ref{fig:ratioarg}. The 78\%
N$_2$ - 21\% O$_2$ - 1\% Ar  air mixture was used for this
measurement.
 The full line is the result of the best fit of Eq. (\ref{eq:rn2}) to the data in the range 2-1000 hPa, with $p'_{\rm{N_2}}$ and $p'_{\rm{O_2}}$ as free parameters. The best fit yielded:
\begin{equation}
p'_{\rm{N_2}}(337) = 103.7 \pm 2.7~ \textrm{hPa},
\label{eq:ppresultn2_1}
\end{equation}
\begin{equation}
p'_{\rm{O_2}}(337) = 3.796 \pm 0.076~ \textrm{hPa},
\label{eq:ppresulto2}
\end{equation}
  where the quoted uncertainties are statistical only, and a \mbox{$\chi^2/\textrm{n.d.f.}= 0.9$}.

 From Eq. (\ref{eq:pprimeair}), a value of  $p'_{air}$ for the  78\% N$_2$ - 21\% O$_2$ - 1\% Ar mixture was estimated:
\begin{equation}
p'_{air}(337) = 15.89 \pm 0.33~ \textrm{hPa},
\label{eq:ppresultair_0}
\end{equation}
where the statistical uncertainty was calculated from the covariance matrix of the fitted $p'_{\rm{N_2}}$ and $p'_{\rm{O_2}}$ parameters.

\begin{figure}
\begin{center}
\includegraphics*[width=14.cm]{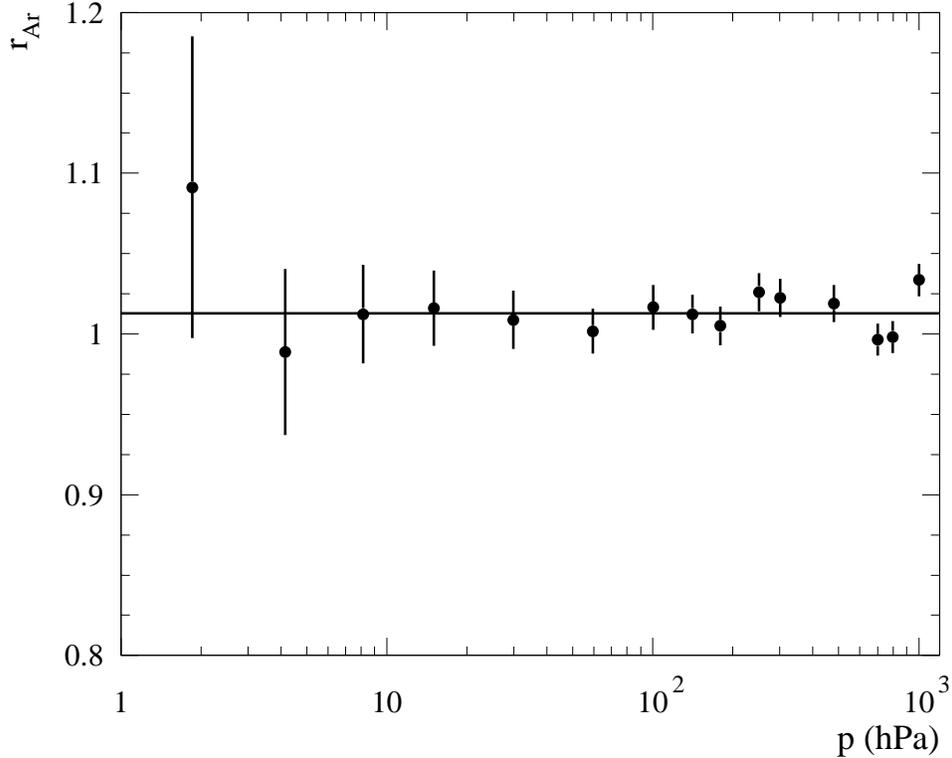}
\end{center}
\caption{Ratio of the fluorescence signal for air-like mixture with
1\% Ar to the mixture without Ar as a function of pressure.}
\label{fig:argovuz}
\end{figure}

Notice that the contribution of argon to the quenching of the 337 nm
band was neglected in Eq. (\ref{eq:rn2}). This is justified by
the comparison of the pressure scan measurements performed with two
air-like mixtures. The ratio $r_{\rm{Ar}}$ of the fluorescence signal
measured  with the 78\% N$_2$ - 21\% O$_2$ - 1\% Ar mixture to the
one with the 79\% N$_2$ - 21\% O$_2$ mixture as a function of
pressure is shown in Fig. \ref{fig:argovuz}. A linear fit to the data yielded 
an intercept of $1.011\pm0.006$ and a slope consistent with zero at the level of $10^{-5}$, with a $\chi^2/\textrm{n.d.f.}= 1.0$. The effect of argon to the quenching of the 337 nm band is thus negligible at all pressures. 

Several checks were performed in order to estimate the systematic uncertainties on the measured $p'$. The stability of the photomultiplier gain was monitored during the measurements by a blue LED. The PMT gain was found to be stable at the 1\% level, with negligible effect on the $p'$ determination.
The range used to fit the fluorescence signal $S_{PMT}$ as a function of the ICT signal $S_{ICT}$ was varied over a wide interval. The lower limit in the fit of the pressure dependence was changed from 2 hPa to 30 hPa. The background was changed between its minimum and maximum measured value. The observed shifts in the $p'$ value due to these variations were taken as estimates of the systematic uncertainties. A systematic uncertainty of 0.1 hPa was assigned from the uncertainty in the measurement of the absolute pressure.
The different contributions to the systematic uncertainty, together with their quadrature sum, are summarized in Table \ref{tab:systpres}. The total uncertainty on the measured $p'$, given by the quadrature sum of the statistical and systematic uncertainties, is also quoted.

\begin{table}[h]
\centering
\begin{tabular}{|c|c|c|c|}
\hline
Source  & $\Delta p'_{\rm{N_2}}$ (hPa) & $\Delta p'_{\rm{O_2}}$ (hPa) & $\Delta p'_{air}$ (hPa)  \\
\hline
slope fit range     & 3.6 & 0.10 & 0.50  \\
\hline
background          & 2.1 & 0.09 & 0.40  \\
\hline
pressure fit range  & 0.3 & 0.01 & 0.03  \\
\hline
absolute pressure   & 0.1 & 0.10 & 0.10  \\
\hline
TOTAL SYST.         & 4.2 & 0.17 & 0.65  \\
\hline
\hline
TOTAL UNCERTAINTY & 5.0 & 0.19 & 0.73  \\
\hline
\end{tabular}
\vskip 0.1truein
\caption{Estimated systematic uncertainties on the quenching reference pressure of the 337 nm fluorescence band. The total uncertainty, given by the the quadrature sum of statistical and systematic uncertainties, is also given in the last line.}
\label{tab:systpres}
\end{table}

The method of the nitrogen to air ratio has the advantage of eliminating the bias due to secondary electrons escaping the field of view of the photomultiplier.
On the other hand, a direct interpretation of the data with Eq. (\ref{eq:signal}) allows a valuable check of our understanding of the experiment.
For this purpose a detailed GEANT4 \cite{geant4} simulation of the experiment was performed. The PENELOPE \cite{penelope} package in GEANT4 was used, which allows the tracking of secondary electrons down to 250 eV. In the simulation, the fluorescence emission was assumed to be proportional to the energy deposited by the particles in the gas. The effective length, which takes into account the field of view losses, was found to be well parameterized by $D(p)=D_{1000}F(p)$ from the Monte Carlo simulations, where $D_{1000}$ is the effective length at 1000 hPa and $F(p)=(p/1000)^{0.027}$. The pressure dependence of the fluorescence signal can thus be fitted with the function:
 \begin{equation}
S_{FL}^{gas}(p) =  C \frac{p}{1+\frac{p}{p'_{gas}(337)}}F(p),
\label{eq:sfit}
\end{equation}
leaving $C$ and $p'_{gas}$ as free parameters.
   A direct fit of the pressure dependence of the 337 nm band fluorescence signal $S^{\rm{N_2}}_{FL}$ in nitrogen with Eq. (\ref{eq:sfit}) yielded:
\begin{equation}
p'_{\rm{N_2}}(337) = 101.0 \pm 0.6~ \textrm{hPa},
\label{eq:ppresultn2_2}
\end{equation}
while a fit of the pressure dependence of the 337 nm fluorescence
signal $S^{air}_{FL}$ in the 78\% N$_2$ - 21\% O$_2$ - 1\% Ar
mixture gave:
\begin{equation}
p'_{air}(337) = 15.34 \pm 0.53~ \textrm{hPa},
\label{eq:ppresultair_1}
\end{equation}
in good agreement with  (\ref{eq:ppresultn2_1}) and (\ref{eq:ppresultair_0}).
To avoid the difficulties of simulations at very low gas densities, we
limited the fit to a minimum pressure of 15 hPa. The fact that two different
ways of taking into account the effect of secondary electrons escaping the
field of view, either by cancelation in the nitrogen to air fluorescence
ratio or by a parameterization of the light losses with simulations, yielded
consistent values of $p'$ strengthen the confidence in our quenching
reference pressure measurements. Also, this consistency shows that the simple ansatz of Eqs. (\ref{eq:fluoreffair})-(\ref{eq:pprimeair}) does indeed provide a good description of our data. 

It should be stressed that the effect of
secondary electrons escaping the detector field of view cannot be neglected
in the measurement of $p'$.  In fact, a fit of the data with $F(p) = 1$ in
Eq. (\ref{eq:sfit}) gave $p'_{\rm{N_2}}(337)=115.5$ hPa and
$p'_{air}(337)=19.95$ hPa, showing a significant bias towards larger $p'$ values.
\section{Pressure dependence of the spectrum bands}
\label{sec:presall}

     The pressure dependence of the air fluorescence spectrum bands has
been measured at the Van de Graaff.  The experimental arrangement and
beam conditions have been described in Sections \ref{sec:expmethod} and
\ref{sec:relintensity}.  The air fluorescence spectrum was measured for
gas pressures in the range from 4 to 1000 hPa, at temperature 293 K.  For each pressure, the
spectrum band intensities relative to the 337 nm band were measured
following the procedure described in Section \ref{sec:relintensity}.
From Eq. (\ref{eq:signal}), the pressure dependence of the
relative intensity of a band of wavelength $\lambda$  is given by:
\begin{equation}
I_{\lambda}(p) =  C
\frac{1+\frac{p}{p'_{air}(337)}}{1+\frac{p}{p'_{air}(\lambda)}}.
\label{eq:pdep2}
\end{equation}

\begin{figure}[htp]
\begin{center}
\includegraphics*[width=15cm]{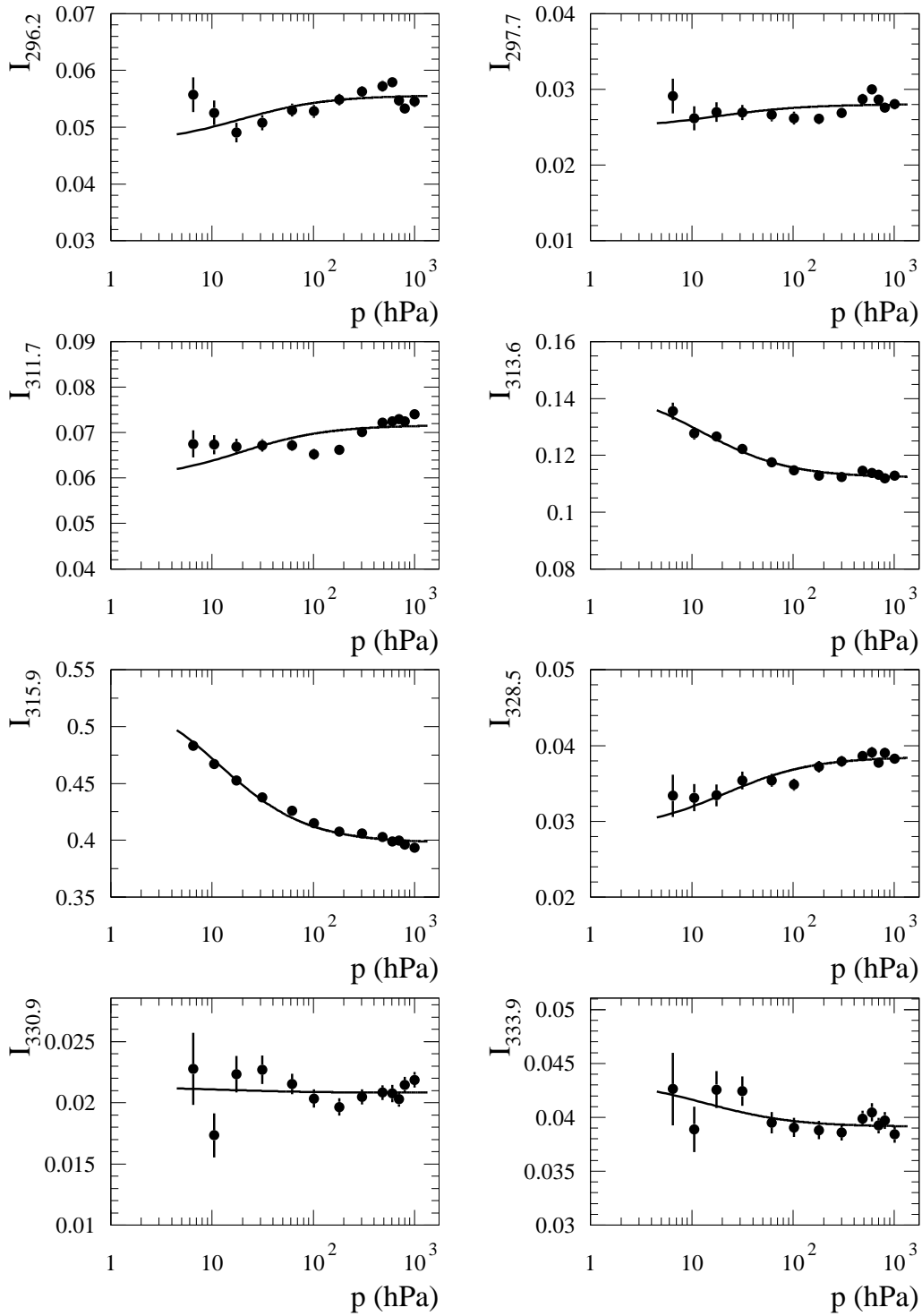}
\end{center}
\caption{Pressure dependence of the air fluorescence bands shown in increasing order from 296.2 nm to 333.9 nm.
Intensities are relative to the 337 nm band. Errors are statistical only. The full line is the result
of a fit described in the text.  }
\label{fig:ratiolambda1}
\end{figure}

\begin{figure}[htp]
\begin{center}
\includegraphics*[width=15cm]{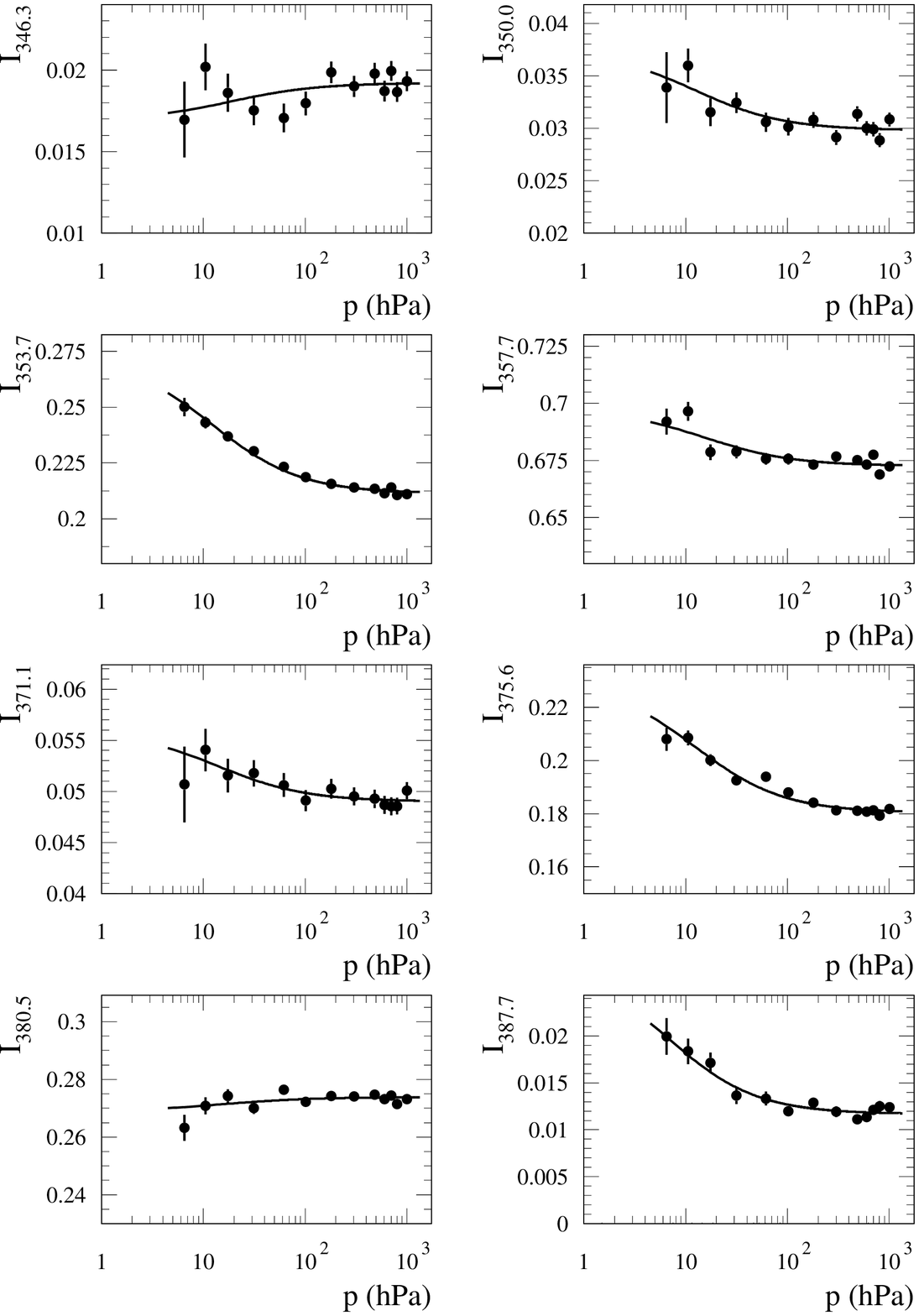}
\end{center}
\caption{Pressure dependence of the air fluorescence bands shown in increasing order from 346.3 nm to 387.7 nm. Intensities are relative to the 337 nm band. Errors are statistical only. The full line is the result of a fit described in the text.}
\label{fig:ratiolambd2}
\end{figure}

\begin{figure}[htp]
\begin{center}
\includegraphics*[width=15cm]{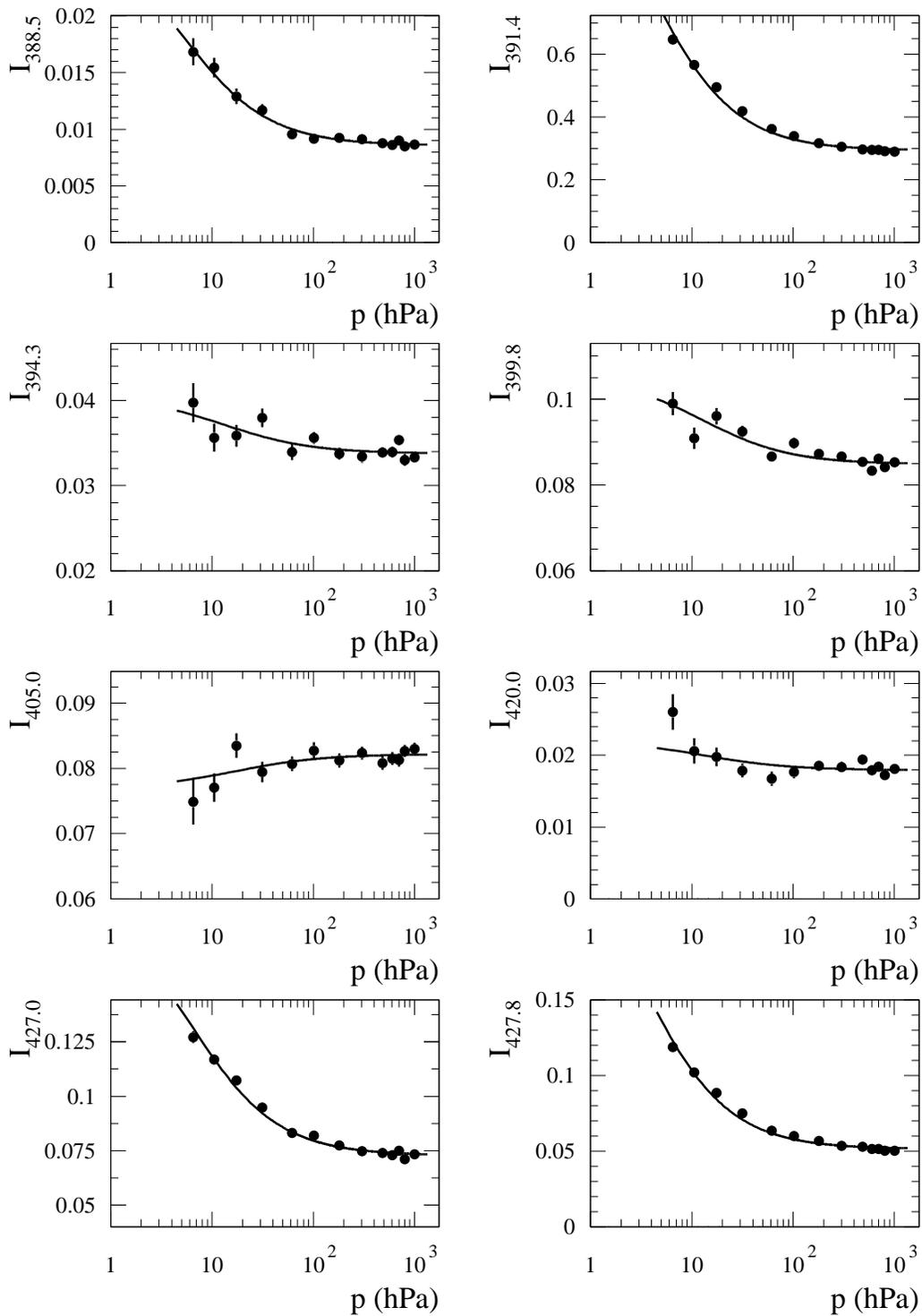}
\end{center}
\caption{Pressure dependence of the air fluorescence bands shown in increasing order from 388.5 nm to 427.8 nm. Intensities are relative to the 337 nm band. Errors are statistical only. The full line is the result of a fit described in the text.}
\label{fig:ratiolambda3}
\end{figure}

The measured $I_{\lambda}(p)$ are shown in Figs. \ref{fig:ratiolambda1}-\ref{fig:ratiolambda3} for one of the ten pressure scans which were performed.
We could not measure the pressure dependence for nine of the bands in Table \ref{tab:relspectrum}, due to their low intensity.
 The full line is the result of the best fit of Eq. (\ref{eq:pdep2}) to the data, leaving $C$ and  $p'_{air}(\lambda)$ as free parameters. The 337 nm quenching reference pressure $p'_{air}(337)$ was fixed to the value of 15.89 hPa obtained in Section \ref{sec:pres337} (Eq. (\ref{eq:ppresultair_0})). 
In principle, the measured $I_{\lambda}(\rm{800~hPa})$ given in Table \ref{tab:relspectrum} could be used as a constraint for the $C$ parameter, therefore reducing the error on  $p'_{air}(\lambda)$. On the other hand, the measured values of  $p'_{air}(\lambda)$ and $I_{\lambda}(\rm{800~hPa})$ would become correlated. 
Leaving $C$ free in the fit also minimizes systematic uncertainties, since the calibration of the relative spectrograph sensitivity does not affect in this case the measured quenching reference pressure. 

Notice that Eq. (\ref{eq:pdep2}) describes a model where the pressure dependence of a spectrum band intensity is given by only one quenching reference pressure.This model may not be true in general because more than one band could be contained in the wavelength interval used for the signal integration. From the quality of the fits, however, it appears that the assumption of a single reference pressure is a good approximation. The largest deviation from the model, $<10$\%, is found for the 311.7 nm band, which is expected to have contributions of similar size from the GH(0,2) and 2P(3,2) bands. A more elaborated model of pressure dependence of the spectrum bands could be used, but since it would not change significantly the application of our measurements to the fluorescence detection of ultra high energy cosmic rays, the simplest approach of Eq. (\ref{eq:pdep2}) was adopted.  

 In order to study possible systematic effects, the measurement was
repeated several times.  Measurements were performed in two separate test
beam periods, under different beam conditions and light collection optics.
Both the 78\% N$_2$ - 21\% O$_2$ - 1\% Ar and the 79\% N$_2$ - 21\% O$_2$
air-like mixtures were used.  All measurements were found to be compatible
within the quoted uncertainties.

 The measured $p'_{air}(\lambda)$, obtained from the weighted average of
ten measurements, are reported in Table \ref{tab:pprime}.  Two uncertainties
are quoted: the first is the r.m.s. of the ten measurements, and the second
comes from changing the value of $p'_{air}(337)$ in the fit by its
total uncertainty $\Delta p'_{air}(337)=\pm 0.73$ hPa (see Table \ref{tab:systpres}).  

\begin{table}[h]
\centering
\begin{tabular}{|c|c|c||c|c|c|}
 \hline
Band & $\lambda$ (nm)  & $p'_{air}(\lambda)$  (hPa) & Band & $\lambda$ (nm)  & $p'_{air}(\lambda)$  (hPa)\\
\hline
2P(0,0) &   337.1  & $15.89 \pm 0.73 $ & 2P(2,0) &   297.7  & $ 17.3 \pm 4.0 \pm 0.8$\\
2P(0,1) &   357.7  & $15.39 \pm 0.25 \pm 0.72$& 2P(2,1) &   313.6  & $ 12.27 \pm 0.78 \pm 0.64$\\
2P(0,2) &   380.5  & $ 16.51 \pm 0.48 \pm 0.72$& 2P(2,2) &   330.9  & $ 16.9  \pm 3.5  \pm 0.76$ \\
2P(0,3) &   405.0  & $ 17.8 \pm 1.5 \pm 0.8$& 2P(2,3) &   350.0  & $ 15.2 \pm 3.7 \pm 0.7 $ \\
\hline
2P(1,0) &   315.9  & $ 11.88 \pm 0.31 \pm 0.62$&  2P(2,4) &   371.1  & $ 14.8 \pm 1.9 \pm 0.7$  \\
2P(1,1) &   333.9  &  $ 15.5 \pm 1.5 \pm 0.7$ & 2P(2,5) &   394.3  & $ 13.7 \pm 3.3 \pm 0.7$\\
2P(1,2) &   353.7  &  $ 12.70 \pm 0.34 \pm 0.64 $&   2P(2,6) &   420.0  &  $ 13.8 \pm 4.0 \pm 0.7 $ \\
2P(1,3) &   375.6  &  $ 12.82 \pm 0.45 \pm 0.62$ & 2P(3,1) &   296.2  & $ 18.5  \pm 5.0  \pm 0.8$\\
2P(1,4) &   399.8  & $ 13.6 \pm 1.1 \pm 0.6$ & 2P(3,2) &   311.7  & $ 18.7  \pm 3.8  \pm 0.8$ \\
2P(1,5) &   427.0  & $  6.38 \pm 0.68 \pm 0.43 $ & 2P(3,3) &   328.5  & $ 20.7 \pm 2.6  \pm 0.8$ \\
\hline
1N(0,0) &   391.4  & $  2.94 \pm 0.58 \pm 0.31$ & GH(0,4) &   346.3  & $  21 \pm 10 \pm 1$ \\
1N(0,1) &   427.8  & $  2.89 \pm 0.64 \pm 0.30$ & GH(0,6) &   387.7  & $  7.6 \pm 1.6 \pm 0.5$\\
\hline
1N(1,1) &   388.5  & $  3.9 \pm 1.7 \pm 0.3$  \\
\cline{1-3}
\end{tabular}
\vskip 0.1truein
\caption{Collisional quenching reference pressures in dry air at 293 K. The quoted uncertainties are explained in the text.}
\label{tab:pprime}.
\end{table}

 Spectrum bands within a system should have the same quenching reference pressure.  This may not be true in our measurements,
since for each band the signal is integrated over a wavelength interval, which can
contain also other bands.  However, most of the bands are well
separated, and contamination is in general small.
Indeed, we observe that most of the bands of a given system have the same
$p'$ within errors. Bands which deviated from this behavior, like the
387.7 and 427.0 nm bands, are significantly contaminated by nearby bands.  This good agreement gives further confidence on the quality of our pressure
dependence measurements. 

\section{Application to the fluorescence detection of ultra high energy cosmic rays}

The number of fluorescence photons emitted at a given stage of a cosmic ray shower development, {\it i.e.} at a given altitude $h$ in the atmosphere, is given by:
\begin{equation}
 N_{\lambda}^{shower}(h) = E_{dep}^{shower}(h) Y_{air}(\lambda,p,T),
\label{eq:nlambdash}
\end{equation}
where $E_{dep}^{shower}(h)$ is the energy deposited by the shower charged particles in the air volume, $p$ and $T$ are the air pressure and temperature at the altitude $h$. Notice that we assumed in Eq. (\ref{eq:nlambdash}) that the fluorescence emission is proportional to the energy deposited in the gas, which has been tested experimentally with good precision \cite{nagano1}\cite{colin}\cite{flash2}.

 Fluorescence telescopes measure $N_{\lambda}^{shower}(h)$ to detect ultra high energy cosmic rays.  If the fluorescence yield $Y_{air}(\lambda,p,T)$
is known, the energy deposited by the cosmic ray shower can be determined.
Integrating along the shower path in the atmosphere will give the electromagnetic energy of the shower.  The fluorescence efficiency must thus be known
over a range of pressure and temperature corresponding to altitudes up to
several tens of km.

It is convenient to express the fluorescence efficiency in terms of the physical quantities measured in this paper, namely the air fluorescence spectrum relative band intensities $I_{\lambda}(p_0,T_0)$ (Section \ref{sec:relintensity}) and  the quenching reference pressures $p'_{air}(\lambda,T_0)$ (Sections \ref{sec:pres337}-\ref{sec:presall}). The air pressure $p_0=800$ hPa and temperature $T_0=293$ K used in the measurements have been introduced in the notation for consistency with what will be used in the following.
 The quenching reference pressure (cf. Eq. (\ref{eq:pprimeair})) can be written in a general way as:
\begin{equation}
 p'_{air}(\lambda,T) = \frac{\sqrt{T}}{A_\lambda H_\lambda(T)},
\label{eq:pprimeairt}
\end{equation}
where $H_\lambda(T)$ has been introduced to take into account a possible temperature dependence of the collisional cross sections.
In terms of the measured quenching reference pressures $ p'_{air}(\lambda,T_0)$, Eq. (\ref{eq:pprimeairt}) becomes:
\begin{equation}
 p'_{air}(\lambda,T) =  p'_{air}(\lambda,T_0) \sqrt{\frac{T}{T_0}}\frac{H_\lambda(T_0)}{H_\lambda(T)}.
\label{eq:pprimeairt0}
\end{equation}

The fluorescence yield can thus be expressed in terms of the measured physical quantities:
\begin{equation}
 Y_{air}(\lambda,p,T)= Y_{air}(337,p_0,T_0) I_{\lambda}(p_0,T_0)\frac{1+\frac{p_0}{p'_{air}(\lambda,T_0)}}{1+\frac{p}{p'_{air}(\lambda,T_0)\sqrt{\frac{T}{T_0}}\frac{H_\lambda(T_0)}{H_\lambda(T)}}} ,
\label{eq:phish}
\end{equation}
This parameterization, together with the values of Tables \ref{tab:relspectrum}, \ref{tab:pprime}, and \ref{tab:pprimeother}, can be used by ultra high energy cosmic rays experiments which employ the fluorescence detection technique. In Table \ref{tab:pprimeother} our best estimates for the unmeasured quenching reference pressure of the nine weak bands not present in Table \ref{tab:pprime} are reported. We assumed in Table \ref{tab:pprimeother} that all the Gaydon-Herman bands have the same $p'$ of the GH(0,4) band. For the unmeasured $p'$ of the 2P(3,i) and 2P(4,i) bands we took the average value of the measured 2P(3,i=1,3) with a large uncertainty. The quenching reference pressure of the 1N(1,2) was taken to be equal to the measured $p'$ of the 1N(1,1).  

\begin{table}[h]
\centering
\begin{tabular}{|c|c|c||c|c|c|}
 \hline
Band & $\lambda$ (nm)  & $p'_{air}(\lambda)$  (hPa)&  Band & $\lambda$ (nm)  & $p'_{air}(\lambda)$  (hPa)\\
\hline
2P(4,4) &   326.8  &  $ 19 \pm 5 \pm 1 $ &GH(6,2)&   302.0  & $21 \pm 10  \pm 1 $ \\
2P(4,7) &   385.8  &  $ 19 \pm 5 \pm 1 $ & GH(5,2)&   308.0  & $21 \pm 10  \pm 1 $\\
\cline{1-3}
2P(3,5) &   367.2  &  $ 19 \pm 5 \pm 1 $  &GH(6,3)&   317.6  & $21 \pm 10  \pm 1 $  \\
2P(3,7) &   414.1  &  $ 19 \pm 5 \pm 1 $  &  GH(0,5)&   366.1  & $21 \pm 10  \pm 1 $\\
\hline
1N(1,2) &   423.6  & $  3.9 \pm 1.7 \pm 0.3$   \\
\cline{1-3}
\end{tabular}
\vskip 0.1truein
\caption{Collisional quenching reference pressures in air at 293 K adopted for the nine weak bands for which $p'$ could not be measured.}
\label{tab:pprimeother}.
\end{table}

Notice that in order to fully determine the fluorescence yield, the absolute fluorescence yield of the 337 nm band, $Y_{air}(337,p_0,T_0)$, and the temperature dependence of the collisional cross sections, $H_\lambda(T)$ must be known.
Measurements of these quantities will be the subject of subsequent papers.

\begin{figure}[t]
\begin{center}
\includegraphics*[width=14.cm]{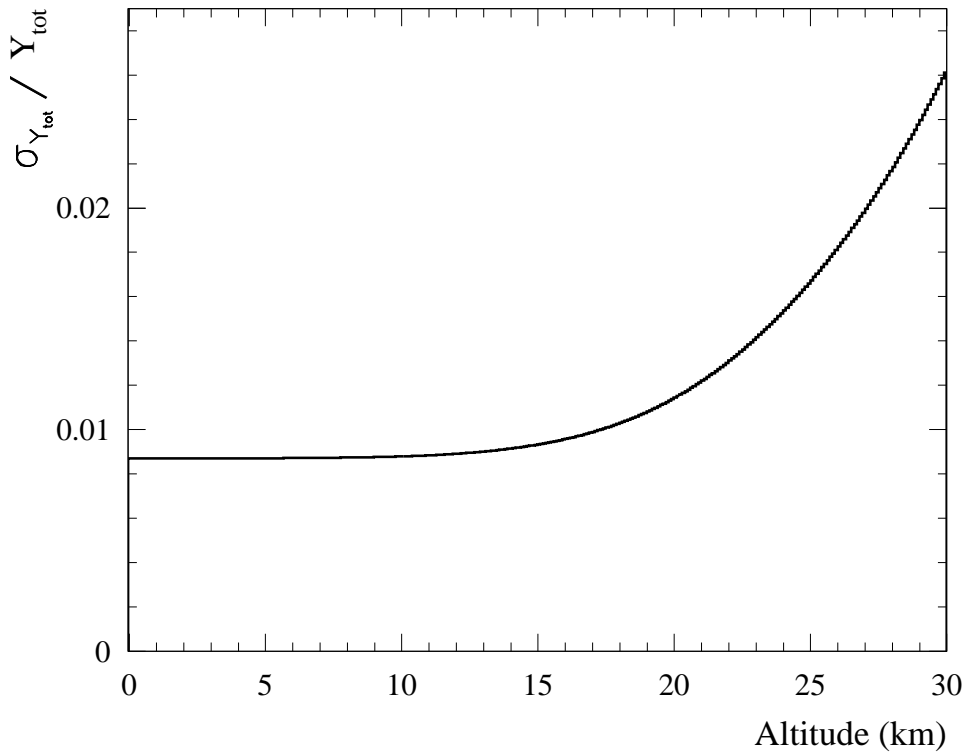}
\end{center}
\caption{Relative uncertainty on the total fluorescence yield as a function of altitude.}
\label{fig:yielderr}
\end{figure}

The uncertainties on the relative band intensities and quenching
reference pressures will result in an uncertainty on the cosmic ray shower
energy. For a proper estimate of this uncertainty, the spectral
sensitivity of the fluorescence detector and the wavelength dependent
atmospheric attenuation must be taken into account because they change the relative weights of the fluorescence spectrum bands. A detailed study of these effects goes beyond the scope of this paper. Nevertherless, an estimate may be obtained from the total fluorescence yield:
\begin{equation}
Y_{tot}(p,T) = \sum_\lambda Y_{air}(\lambda,p,T),
\label{eq:phishnorm}
\end{equation}
where the sum goes over the 34 wavelengths of Table \ref{tab:relspectrum}. 
In Eq.~(\ref{eq:phishnorm}) we assumed  $H_\lambda(T)=1$, {\it i.e.} collisional cross sections have no temperature dependence.
From Eq. (\ref{eq:nlambdash}), the relative uncertainty on the energy deposited by the cosmic ray shower is equal to the relative uncertainty on the total fluorescence yield, $\sigma_{Y_{tot}}/Y_{tot}$.
In Fig. \ref{fig:yielderr}, the relative uncertainty 
on the total fluorescence yield is shown as a function of altitude. 
The U.S. 1976 Standard Atmosphere \cite{stda}  was used to calculate the
pressure and temperature at a given altitude.  
In the calculation of $\sigma_{Y_{tot}}$,
the uncertainties on the relative intensities of the spectrum bands and the quenching reference pressures were treated as uncorrelated, with the exception of the uncertainties related to $p'_{air}$(337) which were fully correlated for all the reference quenching pressures. Ultra high energy cosmic rays deposit most of their energy well below 25 km altitude, which corresponds to a vertical slant depth of only 14 g/cm$^2$. Thus, we may estimate from Fig. \ref{fig:yielderr} that the uncertainties on the measurements reported in this work will result in a relative uncertainty on the cosmic ray shower energy at the level of 1\%. 

\begin{figure}[t]
\begin{center}
\includegraphics*[width=14.cm]{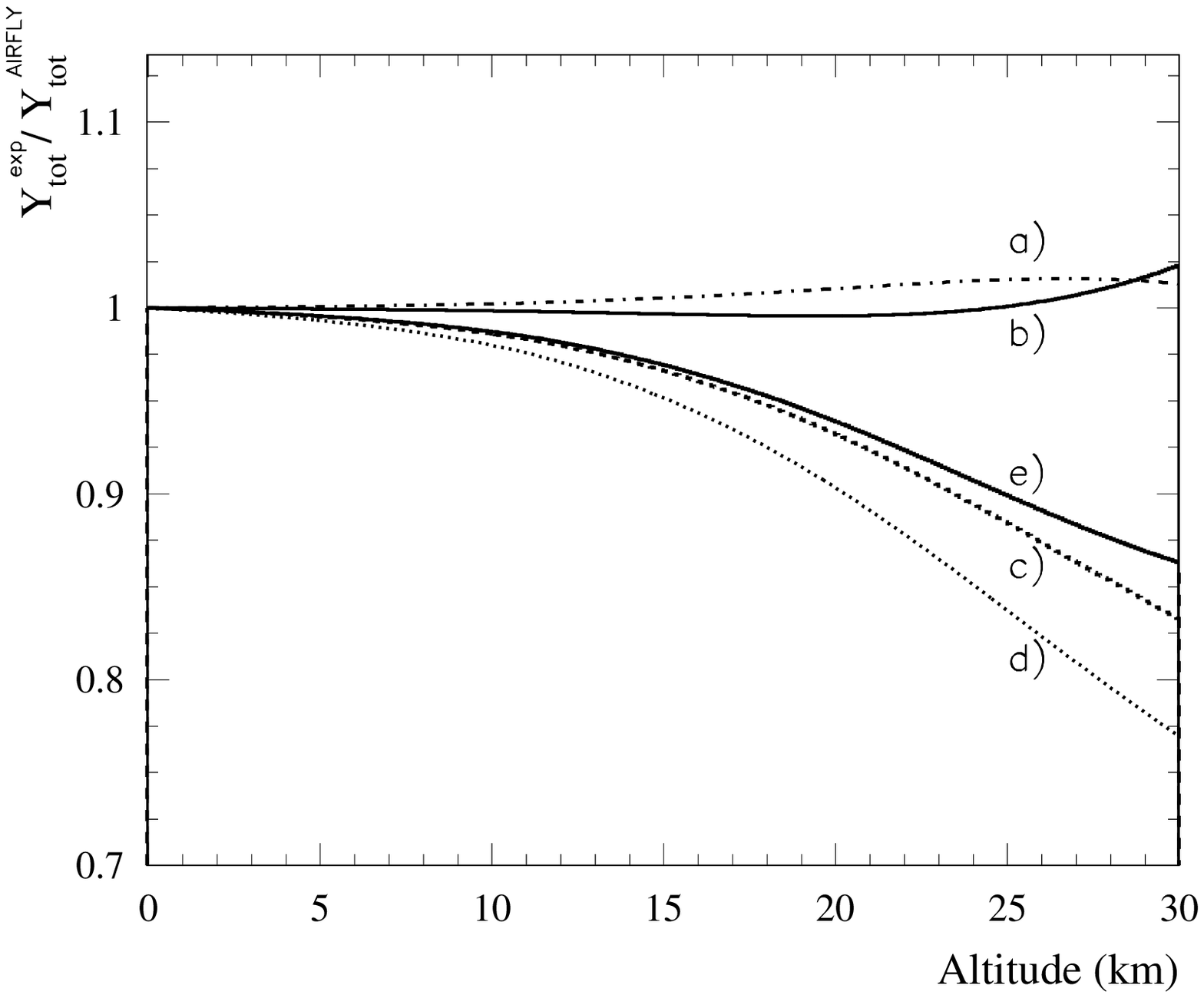}
\end{center}
\caption{Ratio of the total fluorescence yield in the range 300 to 400 nm as measured by other experiments to the one measured by AIRFLY as a function of altitude: a) Belz {\it et al.} b) Bunner c) Nagano {\it et al.} d) Kakimoto {\it et al.} e) Colin {\it et al.} }
\label{fig:yieldexp}
\end{figure}
From our knowledge of the spectrograph's calibration procedures and several checks performed, we estimate that the systematic uncertainty (3.5\%) on the relative intensities of the spectrum bands are largely uncorrelated. Even taking half of this uncertainty as fully correlated when calculating $\sigma_{Y_{tot}}$, the relative uncertainty on the total fluorescence yield would still be very small, of the order of 2\%.

 The AIRFLY fluorescence spectrum is significantly more precise than previous measurements, which had coarser spectrograph resolution \cite{bunner} or made use of narrow band optical filters \cite{kakim} \cite{nagano1}. For the purpose of comparison, the total fluorescence yield in the range 300 to 400 nm can be used, since this quantity was measured also by \cite{belz} and \cite{colin} with a broad band optical filter.

The ratio of the total fluorescence yield in the range 300 to 400 nm as measured by other experiments to the one measured by AIRFLY is shown as a function of altitude in Fig. \ref{fig:yieldexp}. All yields were normalized to have the same value at ground level. Note that optical atmospheric attenuation and scattering effects have not been included in Fig. \ref{fig:yieldexp}. Thus, the impact on the primary cosmic ray energy cannot be directly derived from this figure. Nevertheless, the total yields of Bunner \cite{bunner} and  Belz {\it et al.} \cite{belz} are in very good agreement with AIRFLY. Notice that the measurement of Belz {\it et al.} should not suffer from systematic uncertainties due to secondary electrons, since they measured the pressure dependence of the fluorescence lifetime. Kakimoto {\it et al.} \cite{kakim}, Nagano {\it et al.} \cite{nagano1} and Colin {\it et al.} \cite{colin} present significant deviations, which reflect the systematically higher values of $p'$ measured by these experiments. 

\section{Summary}
We have made a precise measurement of the emission spectrum of nitrogen molecules excited by 3 MeV electrons in dry air. We could measure 34 fluorescence bands in the wavelength range from 284 to 429 nm. The 2P and 1N systems of molecular nitrogen were found to dominate the fluorescence emission. The high resolution spectrograph allowed the identification of a group of weaker bands, which was found to be consistent with the Gaydon-Herman bands. The relative intensities of bands corresponding to transitions from a common upper level were in good agreement with theoretical expectations based on the ratio of Einstein coefficients. 

The pressure dependence of the fluorescence spectrum was also measured from a few hPa up to atmospheric pressure. Particular care was taken to avoid the bias from undetected light due to secondary electrons escaping the detector's field of view at low pressures. For this purpose, the relative intensity of the 337 nm band was measured as a function of pressure in both nitrogen and air, and from the ratio of intensities measured with the two gases, a  measurement of the collisional quenching reference pressure, $p'_{air}$(337), was obtained.
The measurement of the fluorescence spectrum at different pressures allowed the determination of the collisional quenching reference pressures of 24 fluorescence bands, in addition to the 337 nm band. Systematic uncertainties were minimized by taking the intensities relative to the 337 nm band, and by using the value of $p'_{air}$(337) previously measured. Consistent values of the collisional quenching reference pressures were found for bands due to transitions from a common upper level.     

The effect of argon on the fluorescence yield was also investigated, and found to be negligible.

The high resolution spectra recorded by the spectrograph allowed many more closely spaced bands to be separated than in previous experiments. In the ratio of fluorescence intensities, which was extensively used, many systematic uncertainties cancel. Thanks to these improvements in the experimental method, the details and precision of the AIRFLY measurements surpass that of previous experiments.
The application of the AIRFLY results to ultra high energy cosmic ray experiments which employ the fluorescence detection technique was also studied. A parameterization of the fluorescence yield as a function of altitude in terms of the measured relative band intensities and collisional quenching pressures was derived. We estimated that the systematic uncertainty on the cosmic ray shower energy associated with the pressure dependence of the fluorescence spectrum is reduced to a level of 1\% by the AIRFLY results presented in this paper.

\section{Acknowledgements}
The authors wish to thank M. Iannilli, D. Pecchi, L. Sdogati, E. Tusi and G. Vitali of Universit\`{a} di Roma Tor Vergata for the assistance in the construction of the pressure chamber, Ilya A. Shkrob and  R.H. Lowers of the Argonne Chemistry Division for their support and assistance with the Van de Graaff maintenance and operations.
This work was supported by the following agencies and organizations: United States Department of Energy, Office of Science, Offices of High Energy Physics
and Basic Energy Sciences, under contract DE-AC02-06CH11357;
 BMBF grant no. 05 CU5VK1/8 and DFG grant no. KE 1151/1; project GA CR no. 202/05/2470 and project MSMT INGO; Istituto Nazionale di Fisica Nucleare, Italy;  Center of excellence CETEMPS of Universit\`{a} de l'Aquila.



\end{document}